\documentclass[aps,prd,onecolumn,showpacs,showkeys,amsmath,amssymb]{revtex4}
\usepackage{graphicx}
\usepackage{dcolumn}
\usepackage{bm}
\begin{document}
\title{Light Scalar Tetraquark Mesons in the QCD Sum Rule}
%
\author{Hua-Xing Chen$^{1,2}$}
\email{hxchen@rcnp.osaka-u.ac.jp}
\author{Atsushi Hosaka$^{1}$}
\email{hosaka@rcnp.osaka-u.ac.jp}
\author{Shi-Lin Zhu$^{2}$}
\email{zhusl@phy.pku.edu.cn}
\affiliation{$^1$Research Center for Nuclear Physics, Osaka
University, Ibaraki 567--0047, Japan \\$^2$Department of Physics,
Peking University, Beijing 100871, China }
\begin{abstract}
We study the lowest-lying scalar mesons in the QCD sum rule by
considering them as tetraquark states. We find that there are five
independent currents for each state with a certain flavor structure.
By forming linear combinations, we find that some mixed currents
give reliable QCD sum rules. Among various tetraquark currents, we
consider those which are constructed by the diquarks having
anti-symmetric and symmetric flavor structures. That the results of
the QCD sum rule derived from the two types of currents are similar
suggests that the tetraquark states can have a large mixing between
different flavor structures.
\end{abstract}
\maketitle
\pagenumbering{arabic}
%
\section{Introduction}\label{sec_intro}
%

The light scalar mesons $\sigma(600)$, $\kappa(800)$, $a_0(980)$ and
$f_0(980)$ compose a nonet with the mass below 1
GeV~\cite{Yao:2006px,Aitala:2000xu,experiment,Aston:1987ir,Achasov:2000ym,Achasov:2000ku,Akhmetshin:1999di}.
Almost thirty years ago, Jaffe suggested that they can be tetraquark
candidates, which can explain the mass spectrum of the light scalar
mesons and also their decay properties~\cite{Jaffe:1976ig} (See also
Ref.~\cite{Jaffe:2007id} for recent progress).

So far, several different pictures for the scalar mesons have been
proposed. In the conventional quark model, they have a $\bar q q$
configuration of $^3P_0$ whose masses are expected to be larger than
1 GeV due to the $p$-wave orbital excitation~\cite{Close:2002zu}.
Moveover, by a naively counting of the quark mass, the mass ordering
should be $m_\sigma \sim m_{a_0} < m_\kappa < m_{f_0}$. They are
regarded as chiral partners of the Nambu-Goldstone bosons in chiral
models($\pi, K, \eta, \eta^\prime)$~\cite{Hatsuda:1994pi}, and their
masses are expected to be lower than those of the quark model due to
their collective nature. Yet another interesting picture is that
they are tetraquark
states~\cite{Jaffe:2004ph,Brito:2004tv,Maiani:2004uc,Buccella:2006fn,Mathur:2006bs,
Wang:2005xy,Zhang:2006xp,Weinstein:1982gc,Achasov:1997ih}. In
contrast with the $\bar q q$ states, their masses are expected to be
around 0.6 -- 1 GeV with the ordering  of $m_\sigma < m_\kappa <
m_{a_0, f_0}$, consistent with the recent experimental
observations~\cite{Aitala:2000xu,Yao:2006px,experiment}. The
lightness of these states is expected to be explained by the strong
attractive quark correlation in the scalar and isoscalar channel.
There are some lattice studies supporting
this~\cite{Suganuma:2005ds,Liu:2007hm}. Besides their masses, the
decay properties are also interesting and important, and are studied
in many
papers~\cite{Zhou:2004ms,Caprini:2005zr,Guo:2005wp,Guo:2006br,Pennington:2007yt}.

In our previous paper, we found that there are five independent
currents for the tetraquark $u d \bar s \bar s$ of quantum numbers
$J^P=0^+$, and performed a QCD sum rule analysis using both the
single currents and the mixing between two of
them~\cite{Chen:2006hy}. In this paper, we follow the same procedure
and perform the QCD sum rule analysis for the light scalar mesons.
We find once again that there are five independent currents for each
scalar tetraquark state. We perform a reliable QCD sum rule by using
mixed currents, and obtain the masses of the light scalar mesons.
The results are consistent with the experiments. The present
discussion is an extension of  our recent work shortly reported in
Ref.~\cite{Chen:2006zh}.

Unlike $\bar q q$ and $qqq$ currents, tetraquark currents have
complicated structure due to multiquark degrees of freedom. In order
to explain the essential point, it is sufficient to adopt a diquark
construction for tetraquark currents. An alternative method of
mesonic construction is completely equivalent to the
former~\cite{Chen:2006hy}. The tetraquarks contain a diquark and an
antidiquark having either symmetric or antisymmetric flavor
structure. In the flavor SU(3) symmetric limit, they correspond to
$\mathbf{6_f}$ or  $\mathbf{\bar 3_f}$. As we will discuss in the
next section in detail, both diquarks can be used to construct
independent tetraquark currents for scalar mesons. More generally,
there are some independent currents for a given spin with different
flavor structures. This is very much different from the ground state
baryons, where different flavor representations $\mathbf{8}$ and
$\mathbf{10}$ correspond to different spins $\mathbf{1 / 2}$ and
$\mathbf{3 / 2}$, which induce a mass splitting between
$\Delta(1232)$ and $N(939)$.

In this paper, first we construct the tetraquark currents using
diquark and antidiquark fields having the antisymmetric flavor
$\mathbf{\bar 3_f}\otimes\mathbf{3_f}$, which is in accordance with
the expected light scalar nonet. Furthermore, we construct another
set of tetraquark currents by using diquark and antidiquark fields
having the symmetric flavor $\mathbf{6_f} \otimes \mathbf{\bar
6_f}$. We do not, however, consider other possibilities such as
$\mathbf{6_f} \otimes \mathbf{\bar 3_f}$, since they can not produce
tetraquark currents having the scalar quantum numbers (color singlet
and $J^P = 0^+$). Then as we have done
previously~\cite{Chen:2006hy}, we show that there are five
independent currents for both constructions. We will then search
linear combinations of the currents that optimize the QCD sum rule
and reproduce the results compatible with the expected light scalar
mesons. While performing a QCD sum rule analysis, we also find that
the results of the two constructions have some similarities. In
fact, if we work in the $SU(3)_f$ limit, we obtain identical results
for the operator product expansion (OPE).

Since the scalar mesons, especially $\sigma$, decays strongly to two
pseudoscalar mesons, their effects should be significant for
quantitative discussions. The contamination from such two-meson
decay should be removed when performing the QCD sum rule analysis,
which is however a difficult theoretical problem so far.
Nevertheless we consider a phenomenological method by adding another
parameter corresponding to a decay width for the QCD sum rule
analysis.

This paper is organized as follows. In Sec.~\ref{sec_current}, we
establish five independent tetraquark currents of $J^P=0^+$, and
construct mixed currents for $\sigma$, $\kappa$, $a_0$ and $f_0$. In
Sec.~\ref{sec_single}, we perform a QCD sum rule analysis by using
single currents. In Sec.~\ref{sec_mixed}, we perform a QCD sum rule
analysis by using mixed currents. In Sec.~\ref{sec_decay}, we
consider the effect of finite decay width, which is important for
the cases of $\sigma$ and $\kappa$. In Sec.~\ref{sec_meson}, we
perform a QCD sum rule analysis for conventional $\bar q q$ scalar
mesons and compare the result with those of tetraquark sum rule.
Sec.~\ref{sec_summary} is devoted to summary. In
Appendix.~\ref{app_relation}, we study the relations between
$(qq)(\bar q \bar q)$ and $(\bar q q)(\bar q q)$ structures.

%
\section{Tetraquark Currents}\label{sec_current}
%

\begin{figure}[hbt]
\begin{center}
\scalebox{0.9}{\includegraphics{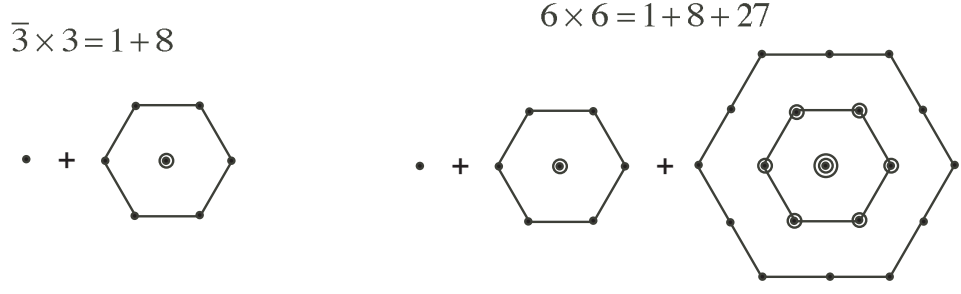}}
 \caption{SU(3) weight
diagrams for tetraquark states of antisymmetric and symmetric
diquarks (antidiquarks).} \label{pic_tetra}
\end{center}
\end{figure}

There are many possibilities to construct tetraquark currents. Let
us classify them first by flavor quantum numbers. In the SU(3)
flavor limit, a diquark or an antidiquark carries the flavor
%
\begin{eqnarray}
\nonumber \mathbf{3_f} \otimes \mathbf{3_f} &=& \mathbf{\bar 3_f}
\oplus \mathbf{6_f}\, , ~~\mbox{or}
\\ \nonumber \mathbf{\bar 3_f} \otimes \mathbf{\bar 3_f} &=&
\mathbf{3_f} \oplus \mathbf{\bar 6_f} \, .
\end{eqnarray}
%
We follow the method in our previous work~\cite{Chen:2006hy}, where
tetraquark currents are formed by a local product of diquark and
antidiquark fields. In order to make a scalar tetraquark current,
the diquark and antidiquark fields should have the same color, spin
and orbital symmetries. Therefore, they must have the same flavor
symmetry, which is either antisymmetric ($\mathbf{\bar 3_f} \otimes
\mathbf{3_f}$) or symmetric ($\mathbf{6_f} \otimes \mathbf{\bar
6_f}$). The possible flavor quantum numbers of the tetraquark states
are then
%
\begin{eqnarray}
\mathbf{\bar 3_f} \otimes \mathbf{3_f} &=&  \mathbf{1_f} \oplus
\mathbf{8_f}\, ,
\nonumber \\
\mathbf{6_f} \otimes \mathbf{\bar 6_f} &=&  \mathbf{1_f}
\oplus  \mathbf{8_f} \oplus \mathbf{27_f} \, ,
\end{eqnarray}
%
where the corresponding weight diagrams are shown in
Fig.~\ref{pic_tetra}. The scalar nonet $\mathbf{1} + \mathbf{8}$ is
therefore included in both representations, independently. For
$\mathbf{\bar 3_f} \times \mathbf{3_f} = \mathbf{1_f} +
\mathbf{8_f}$, $\kappa$ and $a_0$ are the members of $\mathbf{8_f}$
while $\sigma$ and $f_0$ can be either in $\mathbf{1_f}$ or in
isospin $I=0$ component of $\mathbf{8_f}$. Or, they can also mix and
in particular the ideal mixing is achieved by
%
\begin{eqnarray}
|\sigma\rangle &=& \sqrt{\frac{1}{3}}|\mathbf{1_f}\rangle -
\sqrt{\frac{2}{3}} |\mathbf{8_f}, I = 0\rangle\, ,
\nonumber \\
|f_0\rangle&=& \sqrt{\frac{2}{3}} |\mathbf{1_f}\rangle +
\sqrt{\frac{1}{3}} |\mathbf{8_f}, I = 0\rangle\, ,
\end{eqnarray}
%
where only isospin symmetry is respected and the currents are
classified by the number of strange quarks. We can find another set
of linear combinations for the symmetric case. Hence, denoting light
$u, d$ quarks by $q$, $\sigma$ currents are constructed as $qq \bar
q \bar q$, $\kappa$ currents by $qs \bar q \bar q$ and $a_0$ and
$f_0$ currents by $qs \bar q \bar s$. A naive additive quark
counting for this construction is consistent with the observed
masses, $\sigma(600)$, $\kappa(800)$, $a_0(980)$ and $f_0(980)$.
Also, in the QCD sum rule we find that the ideal mixing is needed in
order to reproduce the expected mass pattern of $\sigma$, $\kappa$,
$a_0$ and $f_0$.
%
%

Using the antisymmetric combination for diquark flavor structure, we
arrive at the following five independent currents
%
\begin{eqnarray}
\nonumber\label{define_udud_current} S^\sigma_3 &=& (u_a^T C
\gamma_5 d_b)(\bar{u}_a \gamma_5 C \bar{d}_b^T - \bar{u}_b \gamma_5
C \bar{d}_a^T)\, ,
\\ \nonumber
V^\sigma_3 &=& (u_a^T C \gamma_{\mu} \gamma_5 d_b)(\bar{u}_a
\gamma^{\mu}\gamma_5 C \bar{d}_b^T - \bar{u}_b \gamma^{\mu}\gamma_5
C \bar{d}_a^T)\, ,
\\
T^\sigma_6 &=& (u_a^T C \sigma_{\mu\nu} d_b)(\bar{u}_a
\sigma^{\mu\nu} C \bar{d}_b^T + \bar{u}_b \sigma^{\mu\nu} C
\bar{d}_a^T)\, ,
\\ \nonumber
A^\sigma_6 &=& (u_a^T C \gamma_{\mu} d_b)(\bar{u}_a \gamma^{\mu} C
\bar{d}_b^T + \bar{u}_b \gamma^{\mu} C \bar{d}_a^T)\, ,
\\ \nonumber
P^\sigma_3 &=& (u_a^T C d_b)(\bar{u}_a C \bar{d}_b^T - \bar{u}_b C
\bar{d}_a^T)\, .
\end{eqnarray}
%
where the sum over repeated indices ($\mu$, $\nu, \cdots$ for Dirac,
and $a, b, \cdots$ for color indices) is taken. Either plus or minus
sign in the second parentheses ensures that the diquarks form the
antisymmetric combination in the flavor space. The currents $S$,
$V$, $T$, $A$ and $P$ are constructed by scalar, vector, tensor,
axial-vector, pseudoscalar diquark and antidiquark fields,
respectively. The subscripts $3$ and $6$ show that the diquarks
(antidiquark) are combined into the color representation
$\mathbf{\bar 3_c}$ and $\mathbf{6_c}$ ($\mathbf{3_c}$ or
$\mathbf{\bar 6_c}$), respectively.

We will perform the sum rule analysis using all currents and their
various linear combinations. We will find that the results for
single currents are not always reliable. In fact, we will find a
good sum rule by a linear combination of $A_6^\sigma$ and
$V_3^\sigma$
%
\begin{eqnarray}\label{eq_eta_sigma_1}
\eta^\sigma_1 &=& \cos\theta A^\sigma_6 + \sin\theta V^\sigma_3 \, ,
\end{eqnarray}
%
where $\theta$ is the mixing angle. As we will discuss in
Sec.~\ref{sec_mixed}, the best choice of the mixing angle turns out
to be $\cot\theta = 1/\sqrt{2}$. The mixed currents for $\kappa$,
$a_0$ and $f_0$ can be found in the similar way
%
\begin{eqnarray}\label{eq_eta_kappa_1}
\nonumber \eta^\kappa_1 &=& \cos\theta A^\kappa_6 + \sin\theta
V^\kappa_3 \, ,
\\  \eta^{a_0}_1 &=& \cos\theta A^{a_0}_6 + \sin\theta
V^{a_0}_3 \, ,
\\ \nonumber \eta^{f_0}_1 &=& \cos\theta A^{f_0}_6 + \sin\theta V^{f_0}_3
\, .
\end{eqnarray}
%
where the best choices are still $\cot\theta = 1/\sqrt{2}$.

The QCD sum rule results for $a_0$ and $f_0$ give the same results.
For simplicity, we will use the charged $a_0$ current
%
\begin{eqnarray}\label{eq_eta_a0_1}
\eta^{a_0}_1 &=& \cos\theta A_6^{a_0+} + \sin\theta V^{a_0+}_3
\\ \nonumber &=& \cos\theta (u_a^T C \gamma_{\mu} s_b)(\bar{d}_a
\gamma^{\mu} C \bar{s}_b^T + \bar{d}_b \gamma^{\mu} C \bar{s}_a^T) +
\sin\theta (u_a^T C \gamma_{\mu} \gamma_5 s_b)(\bar{d}_a
\gamma^{\mu}\gamma_5 C \bar{s}_b^T - \bar{d}_b \gamma^{\mu}\gamma_5
C \bar{s}_a^T)\, .
\end{eqnarray}
%

We can also construct the tetraquark currents of $J^P=0^+$ whose
diquark and antidiquark have the symmetric flavor structure. We use
the same superscripts $\sigma$, $\kappa$ and $a_0$ because of the
same quark contents. There are five independent currents
%
\begin{eqnarray}\nonumber
S_6^\sigma &=& q_a^T C \gamma_5 q_b (\bar{q}_a \gamma_5 C
\bar{q}_b^T + \bar{q}_b \gamma_5 C \bar{q}_a^T)\, ,
\\ \nonumber V_6^\sigma &=& q_a^T C \gamma_\mu \gamma_5 q_b (\bar{q}_a \gamma^\mu \gamma_5 C
\bar{q}_b^T + \bar{q}_b \gamma^\mu \gamma_5 C \bar{q}_a^T)\, ,
\\ T_3^\sigma &=& q_a^T C \sigma_{\mu\nu} q_b (\bar{q}_a
\sigma^{\mu\nu} C \bar{q}_b^T - \bar{q}_b \sigma^{\mu\nu} C
\bar{q}_a^T)\, ,
\\ \nonumber A_3^\sigma &=& q_a^T C \gamma_\mu q_b
(\bar{q}_a \gamma^\mu C \bar{q}_b^T - \bar{q}_b \gamma^\mu C
\bar{q}_a^T)\, ,
\\ \nonumber P_6^\sigma
&=& q_a^T C q_b (\bar{q}_a C \bar{q}_b^T + \bar{q}_b C
\bar{q}_a^T)\, .
\end{eqnarray}
%
The quark contents are ${1 \over \sqrt{6}} \big ( \{ u u \} \{ \bar
u \bar u \} - 2 \{ u d \} \{ \bar u \bar d \} + \{ d d  \} \{ \bar d
\bar d \} \big )$ which compose an isoscalar tetraquark. Either plus
or minus sign in the second parentheses ensures that the diquarks
form the symmetric combination in the flavor space. We construct the
similar mixed currents for $\kappa$, $a_0$ and $f_0$
%
\begin{eqnarray}\label{eq_eta_sigma_2}
\nonumber \eta^\sigma_2 &=& \cos\theta A^\sigma_3 + \sin\theta
V^\sigma_6 \, ,
\\ \label{eq_eta_kappa_2} \eta^\kappa_2 &=& \cos\theta A^\kappa_3 + \sin\theta
V^\kappa_6 \, ,
\\ \nonumber \label{eq_eta_a0_2} \eta^{a_0}_2 &=& \cos\theta A^{a_0}_3 + \sin\theta V^{a_0}_6
\, ,
\\ \nonumber \label{eq_eta_a0_2} \eta^{f_0}_2 &=& \cos\theta A^{f_0}_3 + \sin\theta V^{f_0}_6
\, ,
\end{eqnarray}
%
Here the optimal choice of the mixing angle is $\cot\theta=\sqrt{2}$
for $\eta_2^\sigma$ and $\eta_2^{a_0}$, but with a slightly
different value for $\eta_2^\kappa$, which is 1.37.

The currents $\eta_1$ and $\eta_2$ have similar structure. We can
interchange them under the exchange of $\gamma_\mu \leftrightarrow
\gamma_\mu \gamma_5$. We choose the mixing angle $\cot\theta =
1/\sqrt{2}$ for $\eta_1$, which corresponds to $\cot\theta =
\sqrt{2}$ for $\eta_2$.

Concerning linear combinations, we have tested more general cases by
using all five currents. However, we could not find significant
improvements over the present results of using the two currents.

In Table~\ref{table_currents}, we show the diquark properties of ten
single currents. The parity can be obtained by using $P = (-)^L$.
The structures of tetraquark currents are complicated. The flavor
symmetry is not subject to constraints due to the color, spin and
orbital symmetries. If the diquark and antidiquark have the
antisymmetric flavor, they can have both the antisymmetric color
$\mathbf{\bar 3_c} \otimes \mathbf{3_c}$ ($S_3^\sigma$, $V_3^\sigma$
and $P_3^\sigma$) and the symmetric color $\mathbf{6_c} \otimes
\mathbf{\bar 6_c}$ ($T_6^\sigma$ and $A_6^\sigma$); they can have
both the antisymmetric spin $\mathbf{0_S} \otimes \mathbf{0_S}$
($S_3^\sigma$ and $V_3^\sigma$) and the symmetric spin $\mathbf{1_S}
\otimes \mathbf{1_S}$ ($A_6^\sigma$ and $P_3^\sigma$); they can have
both positive parity ($S_3^\sigma$ and $A_6^\sigma$) and negative
parity ($V_3^\sigma$ and $P_3^\sigma$).

The situation is the same for the color, spin and orbital
symmetries. If the diquark and antidiquark have the antisymmetric
color $\mathbf{\bar 3_c} \otimes \mathbf{3_c}$, they can have both
the antisymmetric flavor ($S_3^\sigma$, $V_3^\sigma$ and
$P_3^\sigma$) and the symmetric flavor ($T_3^\sigma$ and
$A_3^\sigma$); they can have both the antisymmetric spin
$\mathbf{0_S} \otimes \mathbf{0_S}$ ($S_3^\sigma$ and $V_3^\sigma$)
and the symmetric spin $\mathbf{1_S} \otimes \mathbf{1_S}$
($A_3^\sigma$ and $P_3^\sigma$); they can have both positive parity
($S_3^\sigma$ and $A_3^\sigma$) and negative parity ($V_3^\sigma$
and $P_3^\sigma$).

We can also construct $(\bar q q)(\bar q q)$ currents. We find that
they are equivalent to the $(qq)(\bar q \bar q)$ currents. We will
explain in detail the relations between $(qq)(\bar q \bar q)$ and
$(\bar q q)(\bar q q)$ structures in the
Appendix.~\ref{app_relation}.

%
\begin{table}
\caption{Diquark properties of single currents.}
\begin{center}
\begin{tabular}{c|ccccc|ccccc}
\hline ($q q$) & $S_3$ & $V_3$ & $T_6$ & $A_6$ & $P_3$ & $S_6$ &
$V_6$ & $T_3$ & $A_3$ & $P_6$
\\ \hline Flavor ($\mathbf{f}$) & $\mathbf{\bar 3}$ & $\mathbf{\bar 3}$ & $\mathbf{\bar 3}$ & $\mathbf{\bar 3}$ & $\mathbf{\bar 3}$
& $\mathbf{6}$ & $\mathbf{6}$ & $\mathbf{6}$ & $\mathbf{6}$ &
$\mathbf{6}$
\\ \hline Color ($\mathbf{c}$) & $\mathbf{\bar 3}$ & $\mathbf{\bar 3}$ & $\mathbf{6}$ & $\mathbf{6}$ & $\mathbf{\bar 3}$
& $\mathbf{6}$ & $\mathbf{6}$ & $\mathbf{\bar 3}$ & $\mathbf{6}$ &
$\mathbf{\bar 3}$
\\ \hline Spin ($\mathbf{S}$) & $\mathbf{0}$ & $\mathbf{0}$ & ($\mathbf{0}$, $\mathbf{1}$) & $\mathbf{1}$ & $\mathbf{1}$ &
$\mathbf{0}$ & $\mathbf{0}$ & ($\mathbf{0}$, $\mathbf{1}$) &
$\mathbf{1}$ & $\mathbf{1}$
\\ \hline Orbit angular momentum ($\mathbf{L}$) & $\mathbf{0}$ & $\mathbf{1}$ & ($\mathbf{1}$, $\mathbf{0}$) & $\mathbf{0}$ &
$\mathbf{1}$ & $\mathbf{0}$ & $\mathbf{1}$ & ($\mathbf{1}$,
$\mathbf{0}$) & $\mathbf{0}$ & $\mathbf{1}$
\\ \hline Total Spin ($\mathbf{J} = \mathbf{S} + \mathbf{L}$) & $\mathbf{0}$ & $\mathbf{1}$ & $\mathbf{1}$ & $\mathbf{1}$ & $\mathbf{0}$ &
$\mathbf{0}$ & $\mathbf{1}$ & $\mathbf{1}$ & $\mathbf{1}$ & $\mathbf{0}$
\\ \hline
\end{tabular}\label{table_currents}
\end{center}
\end{table}
%

%
\section{Analysis of Single Currents}\label{sec_single}
%

In QCD sum rule, we can calculate matrix elements from QCD (OPE) and
relate them to observables by using dispersion relations. Under
suitable assumptions, the QCD sum rule has proven to be a very
powerful and successful non-perturbative method for the past
decades~\cite{Shifman:1978bx,Reinders:1984sr}. Recently, this method
has been applied to study tetraquarks by many
authors~\cite{Narison:2005wc,Bracco:2005kt,Lee:2006vk,Matheus:2007ta,Sugiyama:2007sg}.
In the QCD sum rule analyses, we consider two-point correlation
functions:
%
\begin{equation}
\Pi(q^2)\,\equiv\,i\int d^4x e^{iqx}
\langle0|T\eta(x){\eta^\dagger}(0)|0\rangle \, , \label{eq_pidefine}
\end{equation}
%
where $\eta$ is an interpolating current for the tetraquark. We
compute $\Pi(q^2)$ in the operator product expansion (OPE) of QCD up
to certain order in the expansion, which is then matched with a
hadronic parametrization to extract information of hadron
properties. At the hadron level, we express the correlation function
in the form of the dispersion relation with a spectral function:
%
\begin{equation}
\Pi(p)=\int^\infty_0\frac{\rho(s)}{s-p^2-i\varepsilon}ds \, ,
\label{eq_disper}
\end{equation}
%
where
%
\begin{eqnarray}
\rho(s) & \equiv & \sum_n\delta(s-M^2_n)\langle
0|\eta|n\rangle\langle n|{\eta^\dagger}|0\rangle \ \nonumber\\ &=&
f^2_X\delta(s-M^2_X)+ \rm{higher\,\,states}\, . \label{eq_rho}
\end{eqnarray}
%
For the second equation, as usual, we adopt a parametrization of one
pole dominance for the ground state $X$ and a continuum
contribution. The sum rule analysis is then performed after the
Borel transformation of the two expressions of the correlation
function, (\ref{eq_pidefine}) and (\ref{eq_disper})
%
\begin{equation}
\Pi^{(all)}(M_B^2)\equiv\mathcal{B}_{M_B^2}\Pi(p^2)=\int^\infty_0
e^{-s/M_B^2} \rho(s)ds \, . \label{eq_borel}
\end{equation}
%
Assuming that the contribution from the continuum states can be
approximated well by the spectral density of OPE above a threshold
value $s_0$ (duality), we arrive at the sum rule equation
%
\begin{equation}
\Pi(M_B^2) \equiv f^2_Xe^{-M_X^2/M_B^2} = \int^{s_0}_0
e^{-s/M_B^2}\rho(s)ds \label{eq_fin} \, .
\end{equation}
%
The use of the continuum function of OPE which is the basic
assumption of the duality greatly simplifies the actual sum rule
analyses. Although ambiguities coming from the uncertainties in the
continuum contribution exist~\cite{Lucha:2007pz}, we shall rely on
that assumption as in most of the previous studies. Differentiating
Eq.~(\ref{eq_fin}) with respect to {\Large $\frac{1}{M_B^2}$} and
dividing it by Eq. (\ref{eq_fin}), finally we obtain
%
\begin{equation}
M^2_X=\frac{\int^{s_0}_0 e^{-s/M_B^2}s\rho(s)ds}{\int^{s_0}_0
e^{-s/M_B^2}\rho(s)ds}\, . \label{eq_LSR}
\end{equation}
%

In this section, we show the QCD sum rule analysis of $\kappa$ using
single currents $S^\kappa_3$, $V^\kappa_3$, $T^\kappa_6$,
$A^\kappa_6$ and $P^\kappa_3$. The results for $\sigma$, $a_0$ and
$f_0$ are quite similar. We have performed the OPE calculation up to
dimension eight by using $Mathematica$ with
$FeynCalc$~\cite{FeynCalc}. The results are
%
\begin{eqnarray}
\rho^\kappa_{S3}(s)&=&\frac{s^4} {61440 \pi^6} -\frac{{m_s}^2 s^3}
{3072 \pi^6} + ( \frac{\langle g^2GG \rangle} {6144 \pi^6} -
\frac{{m_s} \langle \bar{q}q \rangle} {192 \pi^4} + \frac{{m_s}
\langle \bar{s}s \rangle} {384 \pi^4}  ) s^2
\nonumber \\
&& + ( - \frac{m_s^2 \langle g^2GG \rangle}{2048\pi^6} - \frac{m_s
\langle g\bar{q} \sigma Gq \rangle}{128\pi^4} + \frac{\langle
\bar{q}q \rangle ^2}{24\pi^2} + \frac{\langle \bar{q}q \rangle
\langle \bar{s}s \rangle }{24\pi^2}) s
\\
\nonumber && - \frac{m_s^2 \langle \bar{q}q \rangle ^2}{12\pi^2} -
\frac{m_s\langle g^2 GG \rangle\langle \bar{q}q \rangle}{768\pi^4} +
\frac{m_s\langle g^2 GG \rangle\langle \bar{s}s \rangle}{1536\pi^4}
+ \frac{\langle \bar{q}q \rangle \langle g\bar{q}\sigma Gq
\rangle}{24\pi^2} + \frac{\langle \bar{s}s \rangle\langle g\bar{q}
\sigma Gq \rangle}{48\pi^2} + \frac{\langle \bar{q}q \rangle\langle
g\bar{s} \sigma Gs \rangle}{48\pi^2}\, ,
\\
\nonumber \rho^\kappa_{V3}(s)&=&\frac{s^4} {15360 \pi^6}
-\frac{{m_s}^2 s^3} {768 \pi^6} +( \frac{\langle g^2GG \rangle}
{3072 \pi^6} + \frac{{m_s} \langle \bar{q}q \rangle} {96 \pi^4} +
\frac{{m_s} \langle \bar{s}s \rangle} {96 \pi^4} ) s^2
\\
&& + ( - \frac{ m_s^2 \langle g^2GG \rangle}{1024\pi^6} + \frac{m_s
\langle g\bar{q} \sigma Gq \rangle}{128\pi^4} -\frac{\langle
\bar{q}q \rangle ^2}{12\pi^2} -\frac{\langle \bar{q}q \rangle
\langle \bar{s}s \rangle }{12\pi^2}) s
\\ \nonumber &&
+\frac{m_s^2 \langle \bar{q}q \rangle ^2}{6\pi^2} - \frac{m_s
\langle g^2GG \rangle \langle \bar{q}q \rangle}{384\pi^4} + \frac{
m_s \langle g^2GG \rangle \langle \bar{s}s \rangle}{768\pi^4}
-\frac{\langle \bar{q}q \rangle \langle g\bar{q}\sigma Gq
\rangle}{12\pi^2} - \frac{\langle \bar{s}s \rangle\langle g\bar{q}
\sigma Gq \rangle}{48\pi^2} - \frac{\langle \bar{q}q \rangle\langle
g\bar{s} \sigma Gs \rangle}{16\pi^2} \, ,
\\
\rho^\kappa_{T6}(s)&=&\frac{s^4} {1280 \pi^6} -\frac{{m_s}^2 s^3}
{64 \pi^6} +( \frac{11 \langle g^2GG \rangle} {768 \pi^6}
+\frac{{m_s} \langle \bar{s}s \rangle} {8 \pi^4}   ) s^2  - \frac{11
m_s^2 \langle g^2GG \rangle}{256\pi^6} s + \frac{11 m_s\langle g^2
GG \rangle\langle \bar{s}s \rangle}{192\pi^4} \, ,
\\
\nonumber \rho^\kappa_{A6}(s)&=&\frac{s^4} {7680 \pi^6}
-\frac{{m_s}^2 s^3} {384 \pi^6} + ( \frac{5 \langle g^2GG \rangle}
{3072 \pi^6} -\frac{{m_s} \langle \bar{q}q \rangle} {48 \pi^4} +
\frac{{m_s} \langle \bar{s}s \rangle} {48 \pi^4} ) s^2
\\
&& + ( - \frac{5 m_s^2 \langle g^2GG \rangle}{1024\pi^6} + \frac{m_s
\langle g\bar{q} \sigma Gq \rangle}{128\pi^4} + \frac{\langle
\bar{q}q \rangle ^2}{6\pi^2} + \frac{\langle \bar{q}q \rangle
\langle \bar{s}s \rangle }{6\pi^2} ) s
\\
\nonumber && - \frac{m_s^2 \langle \bar{q}q \rangle^2} {3\pi^2} -
\frac{m_s\langle g^2 GG \rangle\langle \bar{q}q \rangle}{384\pi^4} +
\frac{ 5 m_s\langle g^2 GG \rangle\langle \bar{s}s
\rangle}{768\pi^4} + \frac{\langle \bar{q}q \rangle \langle
g\bar{q}\sigma Gq \rangle}{6\pi^2} - \frac{\langle \bar{s}s
\rangle\langle g\bar{q} \sigma Gq \rangle}{48\pi^2} + \frac{ 3
\langle \bar{q}q \rangle\langle g\bar{s} \sigma Gs
\rangle}{16\pi^2}\, ,
\\
\nonumber\label{rho_p6} \rho^\kappa_{P3}(s)&=&\frac{s^4} {61440
\pi^6} -\frac{{m_s}^2 s^3} {3072 \pi^6} +( \frac{\langle g^2GG
\rangle} {6144 \pi^6} + \frac{{m_s} \langle \bar{q}q \rangle} {192
\pi^4} + \frac{{m_s} \langle \bar{s}s \rangle} {384 \pi^4}   ) s^2
\\ && + (
- \frac{m_s^2 \langle g^2GG \rangle}{2048\pi^6} +\frac{m_s \langle
g\bar{q} \sigma Gq \rangle}{128\pi^4} -\frac{\langle \bar{q}q
\rangle ^2}{24\pi^2} -\frac{\langle \bar{q}q \rangle \langle
\bar{s}s \rangle}{24\pi^2}) s
\\ \nonumber &&
+\frac{m_s^2 \langle \bar{q}q \rangle ^2}{12\pi^2} +
\frac{m_s\langle g^2 GG \rangle\langle \bar{q}q \rangle}{768\pi^4} +
\frac{m_s\langle g^2 GG \rangle\langle \bar{s}s \rangle}{1536\pi^4}
- \frac{\langle \bar{q}q \rangle\langle g\bar{q} \sigma Gq
\rangle}{24\pi^2} - \frac{\langle \bar{s}s \rangle\langle g\bar{q}
\sigma Gq \rangle}{48\pi^2} - \frac{\langle \bar{q}q \rangle\langle
g\bar{s} \sigma Gs \rangle}{48\pi^2}\, .
\end{eqnarray}
%
In these equations, $q$ represents a $u$ or $d$ quark, and $s$
represents an $s$ quark. $\langle \bar{q}q \rangle$ and $\langle
\bar{s}s \rangle$ are dimension $D=3$ quark condensates; $\langle
g^2 GG \rangle$ is a $D=4$ gluon condensate; $\langle g\bar{q}\sigma
Gq \rangle$ and $\langle g\bar{s}\sigma Gs \rangle$ are $D=5$ mixed
condensates.

For numerical calculations, we use the following values of
condensates~\cite{Yang:1993bp,Narison:2002pw,Gimenez:2005nt,Jamin:2002ev,Ioffe:2002be,Ovchinnikov:1988gk,Yao:2006px}:
%
\begin{eqnarray}
\nonumber &&\langle\bar qq \rangle=-(0.240 \mbox{ GeV})^3\, ,
\\
\nonumber &&\langle\bar ss\rangle=-(0.8\pm 0.1)\times(0.240 \mbox{
GeV})^3\, ,
\\
\nonumber &&\langle g_s^2GG\rangle =(0.48\pm 0.14) \mbox{ GeV}^4\, ,
\\ \nonumber && m_u = 5.3 \mbox{ MeV}\, ,m_d = 9.4 \mbox{ MeV}\, ,
\\
\label{condensates} &&m_s(1\mbox{ GeV})=125 \pm 20 \mbox{ MeV}\, ,
\\
\nonumber && \langle g_s\bar q\sigma G
q\rangle=-M_0^2\times\langle\bar qq\rangle\, ,
\\
\nonumber &&M_0^2=(0.8\pm0.2)\mbox{ GeV}^2\, .
\end{eqnarray}
%
As usual we assume the vacuum saturation for higher dimensional
operators such as $\langle 0 | \bar q q \bar q q |0\rangle \sim
\langle 0 | \bar q q |0\rangle \langle 0|\bar q q |0\rangle$. There
is a minus sign in the definition of the mixed condensate $\langle
g_s\bar q\sigma G q\rangle$, which is different with some other QCD
sum rule calculation. This is just because the definition of
coupling constant $g_s$ is
different~\cite{Yang:1993bp,Hwang:1994vp}.

\begin{figure}[hbt]
\begin{center}
\scalebox{0.9}{\includegraphics{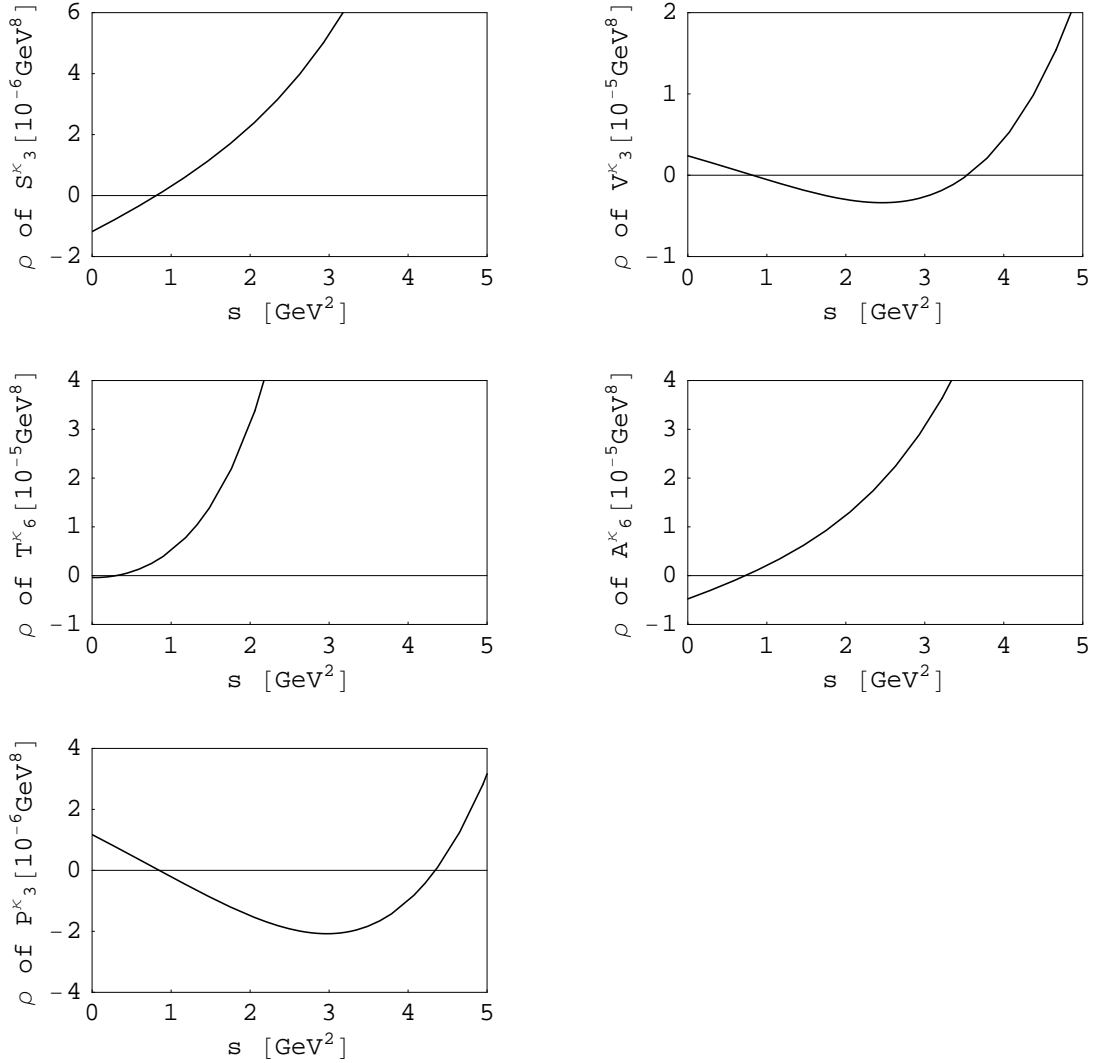}} \caption{Spectral
densities $\rho(s)$ for the currents $S^\kappa_{3}$, $V^\kappa_{3}$,
$T^\kappa_{6}$, $A^\kappa_{6}$ and $P^\kappa_{3}$ as functions of
$s$, in units of $\mbox{GeV}^{8}$.} \label{pic_single}
\end{center}
\end{figure}

For each single current, we have tested the QCD sum rule analysis,
but the result is not good just as in our previous
paper~\cite{Chen:2006hy}. The spectral densities are shown in
Fig.~\ref{pic_single} as functions of the energy square $s$. Due to
the insufficient convergence of the OPE, the positivity of $\rho(s)$
may not be realized. We find that two functions of $S^\kappa_3$ and
$A^\kappa_6$ currents show such a bad behavior that $\rho(s)$
becomes negative in the region of $s = 0 \sim 1$ GeV$^2$, and the
QCD sum rule for these two single currents are not reliable.
%
%

The convergence of the OPE is another important issue. We show the
Borel transformed correlation functions for positive case of
$V^\kappa_3$, $T^\kappa_6$ and $P^\kappa_3$ with numerical
coefficients:
%
\begin{eqnarray}\nonumber\label{rho_pol_qq}
\Pi_{V3}^{\kappa(all)} &=& 1.6 \times 10^{-6} M_B^{10} - 1.3 \times
10^{-7} M_B^8 - 3.5 \times 10^{-6} M_B^6 - 2.8 \times 10^{-6} M_B^4
+ 2.4 \times 10^{-6} M_B^2\, ,
\\ \nonumber \Pi_{T6}^{\kappa(all)} &=& 2.0 \times 10^{-5} M_B^{10} - 1.5 \times 10^{-6}
M_B^8 + 1.1 \times 10^{-5} M_B^6 - 3.3 \times 10^{-7} M_B^4 - 3.9
\times 10^{-7} M_B^2\, ,
\\ \nonumber \Pi_{P3}^{\kappa(all)} &=& 4.1 \times 10^{-7} M_B^{10} - 3.2 \times 10^{-8}
M_B^8 - 9.8 \times 10^{-8} M_B^6 - 1.4 \times 10^{-6} M_B^4 + 1.2
\times 10^{-6} M_B^2\, .
\\
\end{eqnarray}
%
From these expressions, we observe that the convergence of the
currents $V^\kappa_3$ and $P^\kappa_3$ is not very good at a typical
energy scale $M_B \sim 1$ GeV. We have also calculated the pole
contribution which is defined as
%
\begin{equation}\label{eq_pole}
\mbox{Pole contribution} \equiv \frac{ \int^{s_0}_0 e^{-s/M_B^2}
\rho(s)ds }{\int^\infty_0 e^{-s/M_B^2} \rho(s)ds}\, ,
\end{equation}
%
However, due to the negative part of the spectral densities, the
pole contribution is not well defined. Take the current $P^\kappa_3$
as an example, when we choose $s_0 = 1$ GeV$^2$ and $M_B = 0.5$ GeV,
the pole contribution is $101 \%$, which is larger than $100 \%$,
and does not make sense. The pole contribution is $26 \%$ for the
current $T_6^\kappa$, when we choose $s_0 = 1$ GeV$^2$ and $M_B =
0.6$ GeV.

Summarizing the QCD sum rule analysis for the single currents,
including both the $(qq)(\bar q \bar q)$ currents and $(\bar q
q)(\bar q q)$, we found that $T_6^\kappa$ gives the best QCD sum
rule, which however is not yet good enough for quantitative
discussion. In order to improve the sum rule, we move on to study
their linear combinations, which are the mixed currents.

%
\section{Analysis of Mixed Currents}\label{sec_mixed}
%

We have performed the OPE calculation for the mixed currents
$\eta_1$ and $\eta_2$ up to dimension eight, which contains the
four-quark condensates. The $u$ and $d$ quark masses are considered
in the case of the $\sigma$ meson, and neglected in other cases.
%
\begin{eqnarray}\label{rho_sigma1}
\rho^\sigma_1(s) &=& {1 \over 11520 \pi^6} s^4 - {m_u^2 + m_d^2
\over 288 \pi^6} s^3 + \big ( { 6 \sqrt{2} + 7 \over 9216 \pi^6}
\langle g^2 G G \rangle + {(m_u + m_d) \langle \bar q q \rangle
\over 36 \pi^4} \big ) s^2
\\ \nonumber &&
+ \big ( - {6 \sqrt{2} + 7 \over 1536 \pi^6 } ( m_u^2 + m_d^2 )
\langle g^2 G G \rangle + { m_u m_d \langle g^2 G G \rangle \over
512 \pi^6 } - { (m_u^3 + 4 m_u^2 m_d + 4 m_u m_d^2 + m_d^3 ) \langle
\bar q q \rangle \over 6 \pi^4 } \big ) s
\\ \nonumber && + { (5 m_u^2 + 20 m_u m_d + 5 m_d^2 ) \langle \bar q q
\rangle^2 \over 9 \pi^2 } + { 6 \sqrt{2} + 1 \over 1152 \pi^4 } (
m_u + m_d ) \langle g^2 G G \rangle \langle \bar q q \rangle - { (
m_u^2 m_d + m_u m_d^2 ) \langle \bar q \sigma G q \rangle \over 6
\pi^4 } \, ,
\label{rho_sigma2}
\\ \rho^\sigma_2(s) &=& {1 \over 11520 \pi^6} s^4 - {m_u^2 + m_d^2 \over 288 \pi^6}
s^3 + \big ( { 6 \sqrt{2} + 7 \over 9216 \pi^6} \langle g^2 G G
\rangle + {(m_u + m_d) \langle \bar q q \rangle \over 36 \pi^4} \big
) s^2
\\ \nonumber &&
+ \big ( - {4 \sqrt{2} + 5 \over 1024 \pi^6 } ( m_u^2 + m_d^2 )
\langle g^2 G G \rangle - { m_u m_d \langle g^2 G G \rangle \over
768 \pi^6 } - { ( 7 m_u^3 + 8 m_u^2 m_d + 8 m_u m_d^2 + 7 m_d^3 )
\langle \bar q q \rangle \over 18 \pi^4 } \big ) s
\\ \nonumber && + { (25 m_u^2 + 40 m_u m_d + 25 m_d^2 ) \langle \bar q q
\rangle^2 \over 27 \pi^2 } + { 6 \sqrt{2} + 13 \over 1152 \pi^4 } (
m_u + m_d ) \langle g^2 G G \rangle \langle \bar q q \rangle \\
\nonumber && - { ( m_u^3 + 2 m_u^2 m_d + 2 m_u m_d^2 + m_d^3 )
\langle \bar q \sigma G q \rangle \over 18 \pi^4 } \, ,
\label{rho_kappa1}
\\ \nonumber \rho^\kappa_1(s) &=& {1 \over 11520 \pi^6} s^4 - {m_s^2 \over 572 \pi^6}  s^3 +
\big ( { 6 \sqrt{2} + 7 \over 9216 \pi^6} \langle g^2 G G \rangle +
{m_s \langle \bar s s \rangle \over 72 \pi^4} \big ) s^2 + \big ( -
{6 \sqrt{2} + 7 \over 3072 \pi^6 } m_s^2 \langle g^2 G G \rangle + {
m_s \langle \bar q \sigma G q \rangle \over 128 \pi^4 } \big ) s
\\ && - { m_s \langle g^2 G G \rangle \langle \bar q q \rangle
\over 384 \pi^4} - { \langle \bar s s \rangle \langle \bar q \sigma
G q \rangle \over 48 \pi^2 } + { \langle \bar q q \rangle \langle
\bar s \sigma G s \rangle \over 48 \pi^2 } + { 6 \sqrt{2} + 7 \over
2304 \pi^4 } m_s \langle g^2 G G \rangle \langle \bar s s \rangle \,
,
\label{rho_kappa2}
\\ \nonumber \rho^\kappa_2(s) &=& {1 \over 11520 \pi^6} s^4 - {m_s^2 \over 572 \pi^6}  s^3 +
\big ( { 6 \sqrt{2} + 7 \over 9216 \pi^6} \langle g^2 G G \rangle +
{m_s \langle \bar s s \rangle \over 72 \pi^4} \big ) s^2 + \big ( -
{6 \sqrt{2} + 7 \over 3072 \pi^6 } m_s^2  \langle g^2 G G \rangle -
{ m_s \langle \bar q \sigma G q \rangle \over 128 \pi^4 } \big ) s
\\ && + { m_s \langle g^2 G G \rangle \langle \bar q q \rangle
\over 384 \pi^4} + { \langle \bar s s \rangle \langle \bar q \sigma
G q \rangle \over 48 \pi^2 } - { \langle \bar q q \rangle \langle
\bar s \sigma G s \rangle \over 48 \pi^2 } + { 6 \sqrt{2} + 7 \over
2304 \pi^4 } m_s \langle g^2 G G \rangle \langle \bar s s \rangle \,
,
\label{rho_a0f01}
\\ \nonumber \rho^{a_0}_1(s) &=& {1 \over 11520 \pi^6} s^4 - {m_s^2 \over 288 \pi^6}  s^3 +
\big ( { 6 \sqrt{2} + 7 \over 9216 \pi^6} \langle g^2 G G \rangle +
{m_s \langle \bar s s \rangle \over 36 \pi^4} \big ) s^2 + \big ( -
{6 \sqrt{2} + 7 \over 1536 \pi^6 } m_s^2  \langle g^2 G G \rangle -
{ m_s^3 \langle \bar s s \rangle \over 6 \pi^4 } \big ) s \\ && - {
m_s \langle g^2 G G \rangle \langle \bar q q \rangle \over 192
\pi^4} + { 4 m_s^2 \langle \bar q q \rangle^2 \over 9 \pi^2 }  + { 4
m_s^2 \langle \bar s s \rangle^2 \over 9 \pi^2 } + { 6 \sqrt{2} + 7
\over 1152 \pi^4 } m_s \langle g^2 G G \rangle \langle \bar s s
\rangle \, ,
\\ \nonumber \rho^{a_0}_2(s) &=& {1 \over 11520 \pi^6} s^4 - {m_s^2 \over 288 \pi^6}  s^3 +
\big ( { 6 \sqrt{2} + 7 \over 9216 \pi^6} \langle g^2 G G \rangle +
{m_s \langle \bar s s \rangle \over 36 \pi^4} \big ) s^2 + \big ( -
{6 \sqrt{2} + 7 \over 1536 \pi^6 } m_s^2  \langle g^2 G G \rangle -
{ m_s^3 \langle \bar s s \rangle \over 6 \pi^4 } \big ) s
\\ && + { m_s \langle g^2 G G \rangle \langle \bar q q \rangle \over 192 \pi^4}
+ { 4 m_s^2 \langle \bar q q \rangle^2 \over 9 \pi^2 }  + { 4 m_s^2
\langle \bar s s \rangle^2 \over 9 \pi^2 } + { 6 \sqrt{2} + 7 \over
1152 \pi^4 } m_s \langle g^2 G G \rangle \langle \bar s s \rangle\,
.\label{rho_a0f02}
\end{eqnarray}
%
For $\sigma$, terms containing $u, d$ quark masses $m_q$ are small.
For instance, the term of $m_q \langle \bar q q \rangle$ of
dimension four is about ten times smaller than the other term of
$\langle g^2 G G \rangle$. For $\kappa$, $a_0$ and $f_0$, the terms
containing strangle quark mass are important but those containing
$u$ and $d$ quark masses are negligibly small. Therefore, we have
ignored them in our sum rule analysis.

To obtain a reliable a QCD sum rule, the mixed currents $\eta_1$ and
$\eta_2$ are chosen with the following requirements:
%
\begin{enumerate}

\item
The OPE has a good convergence as going to terms of higher
dimensional operators. This can be examined by the following
numerical Borel transformed correlation functions, which have a good
convergence
%
\begin{eqnarray}
\nonumber \Pi^{\sigma(all)}_{1}(M_B^2) &=& 2.2 \times 10^{-6}
M_B^{10} - 2.5 \times 10^{-9} M_B^{8} + 1.5 \times 10^{-6} M_B^{6} -
4.4 \times 10^{-10} M_B^{4} - 4.8 \times 10^{-9} M_B^{2}\, ,
\\ \nonumber \Pi^{\sigma(all)}_{2}(M_B^2) &=& 2.2 \times 10^{-6} M_B^{10}
- 2.5 \times 10^{-9} M_B^{8} + 1.5 \times 10^{-6} M_B^{6} - 5.3
\times 10^{-10} M_B^{4} - 1.5 \times 10^{-8} M_B^{2}\, ,
\\ \nonumber \Pi^{\kappa(all)}_{1}(M_B^2) &=& 2.2
\times 10^{-6} M_B^{10} - 1.7 \times 10^{-7} M_B^{8} + 1.3 \times
10^{-6} M_B^{6} + 7.2 \times 10^{-8} M_B^{4} - 2.3 \times 10^{-8}
M_B^{2}\, ,
\\ \nonumber \Pi^{\kappa(all)}_{2}(M_B^2) &=& 2.2
\times 10^{-6} M_B^{10} - 1.7 \times 10^{-7} M_B^{8} + 1.3 \times
10^{-6} M_B^{6} - 2.8 \times 10^{-7} M_B^{4} + 3.4 \times 10^{-8}
M_B^{2}\, ,
\\ \nonumber \Pi^{{a_0}(all)}_{1}(M_B^2) &=& 2.2
\times 10^{-6} M_B^{10} - 3.4 \times 10^{-7} M_B^{8} + 8.8 \times
10^{-7} M_B^{6} - 4.1 \times 10^{-8} M_B^{4} + 1.1 \times 10^{-7}
M_B^{2}\, ,
\\ \nonumber \Pi^{{a_0}(all)}_{2}(M_B^2) &=& 2.2
\times 10^{-6} M_B^{10} - 3.4 \times 10^{-7} M_B^{8} + 8.8 \times
10^{-7} M_B^{6} - 4.1 \times 10^{-8} M_B^{4} + 2.3 \times 10^{-8}
M_B^{2}\, .
\end{eqnarray}
%
It is interesting to observe that the correlation functions of
$\sigma$ have the most rapid convergence, justifying the use of a
smaller Borel mass $M_B$ than the other cases of $\kappa$, $a_0$ and
$f_0$.

\item
The spectral densities $\rho(s)$ become positive for almost all
energy values, as shown in Fig.~\ref{pic_mixing}. This can be
examined for all the mixed currents except $\eta^\kappa_2$.
Therefore, we need to change the mixing angle of $\eta^\kappa_2$ a
little, which is from $\sqrt{2}$ to 1.37.
%
\begin{figure}[hbt]
\begin{center}
\scalebox{0.9}{\includegraphics{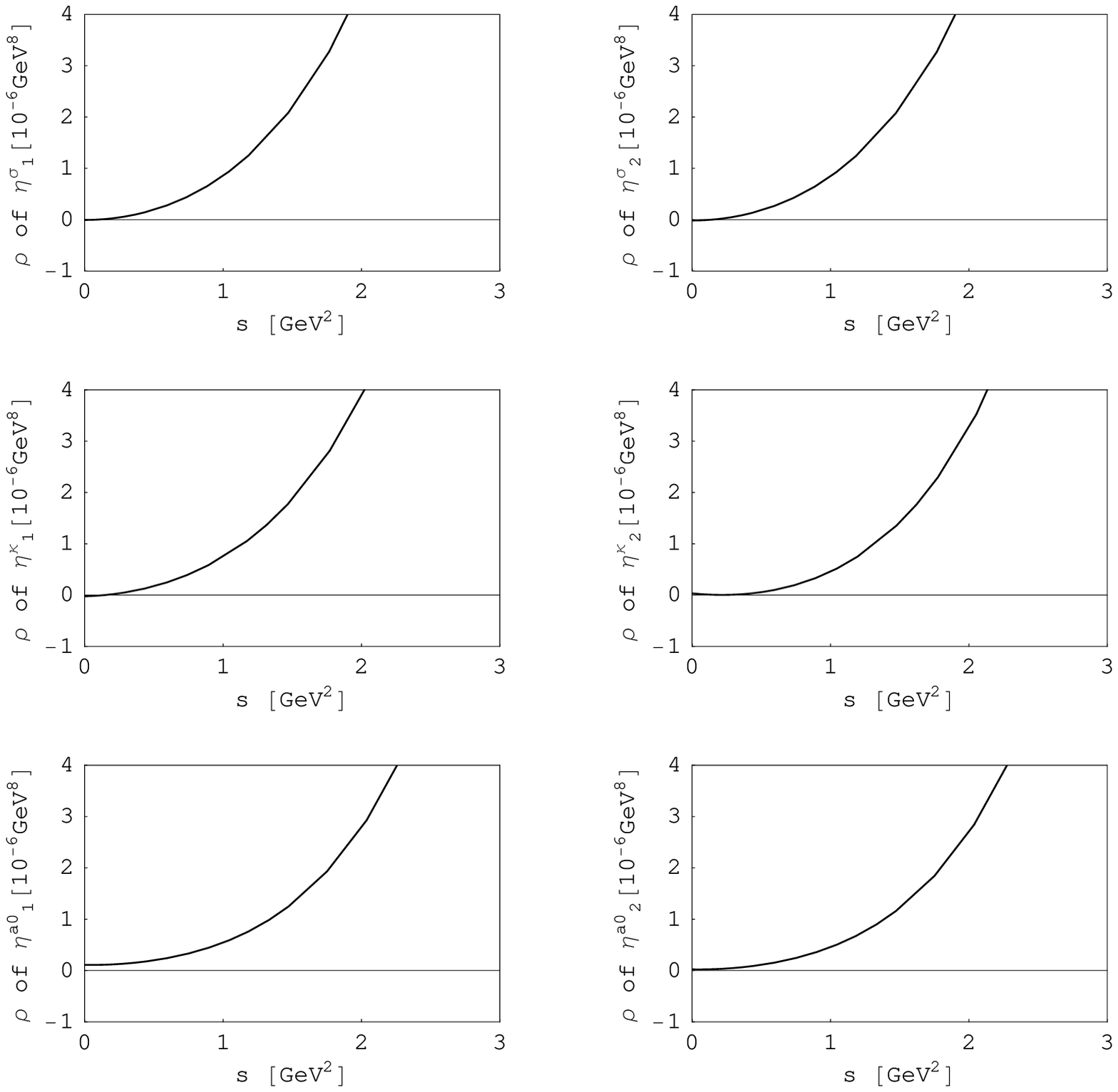}} \caption{Spectral
densities $\rho(s)$ for $\eta^\sigma_1$, $\eta^\sigma_2$,
$\eta^\kappa_1$, $\eta^\kappa_2$, $\eta^{a0,f0}_1$ and
$\eta^{a0,f0}_2$, as functions of $s$, in units of
$\mbox{GeV}^{8}$.} \label{pic_mixing}
\end{center}
\end{figure}
%

\item
Pole contribution is sufficiently large. By choosing suitable Borel
mass $M_B$ and threshold value $s_0$, this can be satisfied. The
Borel transformed correlation functions are written as power series
of the Borel mass $M_B$. Since the Borel transformation suppresses
the contributions from $s > M_B$, smaller values are preferred to
suppress the continuum contributions also. However, for smaller
$M_B$ convergence of the OPE becomes worse. Therefore, we should
find an optimal $M_B$ preferably in a small value region. We have
found that the minima of such a region are 0.5 GeV for $\sigma$, 0.6
GeV for $\kappa$ and 0.8 GeV for $a_0$ and $f_0$, where the pole
contributions reach around 50 \% for $\kappa$, $a_0$ and $f_0$, and
is an acceptable amount for $\sigma$, as shown in
Table~\ref{table_pole}. The pole contribution for the mixed current
$\eta^\kappa_1$ is improved as compared with the single current
$T^\kappa_6$.
%
\begin{table}[hbt]
\caption{Pole contributions of various currents.}
\begin{center}
\begin{tabular}{c|cc|cc|cc}
\hline & $\eta^\sigma_1$ & $\eta^\sigma_2$ & $\eta^\kappa_1$ &
$\eta^\kappa_2$ & $\eta^{a_0}_1$ & $\eta^{a_0}_2$
\\ \hline $M_B$ (GeV) & 0.5 & 0.5  & 0.6 & 0.6 & 0.8 & 0.8
\\ \hline $\sqrt{s_0}$ (GeV) & 0.7 & 0.7 & 1  & 1 & 1.3 & 1.3
\\ \hline Pole (\%) & 28 & 21 & 45 & 36 & 40 & 32
\\ \hline
\end{tabular}\label{table_pole}
\end{center}
\end{table}
%

\end{enumerate}
%

In the $SU(3)_f$ limit, we could find that the differences between
$\rho_1$ and $\rho_2$ vanish:
%
\begin{eqnarray}
\nonumber \rho^\sigma_1(s) - \rho^\sigma_2(s) &=& { (m_u^2 + m_d^2 )
\langle g^2 G G \rangle \over 3072 \pi^6 }s + { 5 m_u m_d \langle
g^2 G G \rangle \over 1536 \pi^6 }s + { (2 m_u^3 - 2 m_u^2 m_d - 2
m_u m_d^2 + 2 m_d^3 ) \langle \bar q q \rangle \over 9 \pi^4 } s
\\ \nonumber && + { ( -10 m_u^2 + 20 m_u m_d - 10 m_d^2 ) \langle \bar q q
\rangle^2 \over 27 \pi^2 } - { ( m_u + m_d ) \langle g^2 G G \rangle
\langle \bar q q \rangle \over 96 \pi^4 }  \\ \nonumber && + { (
m_u^3 - m_u^2 m_d - m_u m_d^2 + m_d^3 ) \langle \bar q \sigma G q
\rangle \over 18 \pi^4 } \, ,
\\ \rho^\kappa_1(s) - \rho^\kappa_2(s) &=& { m_s \langle \bar q \sigma G q \rangle \over 64 \pi^4 } s -
{ m_s \langle g^2 G G \rangle \langle \bar q q \rangle \over 192
\pi^4 } - { \langle \bar s s \rangle \langle \bar q \sigma G q
\rangle \over 24 \pi^2 } + { \langle \bar q q \rangle \langle \bar s
\sigma G s \rangle \over 24 \pi^2 } \, ,
\\ \nonumber \rho^{a_0}_1(s) - \rho^{a_0}_2(s) &=& -
{ m_s \langle g^2 G G \rangle \langle \bar q q \rangle \over 96
\pi^4 }\, .
\end{eqnarray}
%

From Eqs.~(\ref{rho_sigma1}) - (\ref{rho_a0f02}), we find that the
gluon condensates are quite important.  In the chiral limit where
all quark masses vanish, the masses of the scalar mesons are
dictated only by the gluon condensate. Due to the small $u$ and $d$
quark masses, the mass of the $\sigma$ is dominated by the gluon
condensate. For other masses, however, other condensates with $m_s$
also play a significant role. As quarks (in particular strange
quark) become massive, the degeneracy resolves. We have also tested
the case of the SU(3) limit but with the average quark mass, $m_q
\sim 50$ MeV, and with average condensates. Then the mass of the
scalar mesons turns out to be about 0.8--0.9 GeV.

If the location of a physical state is well separated from the
threshold $s_0$, slight change in $s_0$ should not affect much on
the observables (mass) of the state. Hence we have searched the
region where the tetraquark mass varies significantly less than the
change in $\sqrt{s_0}$. We have found such regions for $s_0$ at
around 1 GeV$^2$ from the minimum for $\sigma$ $s_0({\rm min}) \sim
0.5$ GeV${^2}$, for $\kappa$ $s_0({\rm min}) \sim 1$ GeV${^2}$ and
for $a_0$ and $f_0$ $s_0({\rm min}) \sim 1.7$ GeV${^2}$, and up to
about 1 GeV$^2$ higher.

\begin{figure}[hbt]
\begin{center}
\scalebox{0.8}{\includegraphics{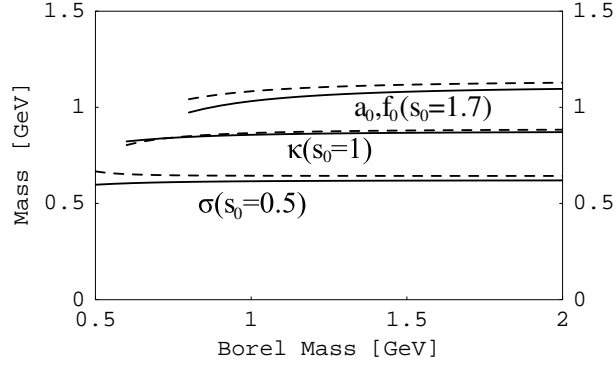}} \caption{Masses of the
$\sigma$, $\kappa$, $a_0$ and $f_0$ as tetraquark states calculated
by the mixed currents $\eta_1$ (solid line) and $\eta_2$ (dashed
line), as functions of the Borel mass $M_B$.} \label{pic_mass}
\end{center}
\end{figure}

After careful test of the sum rule for a wide range of parameter
values of $M_B$ and $s_0$, we have found reliable sum rules, which
are shown in Table~\ref{table_mass}. It is interesting to observe
that the masses appear roughly in the order of the number of strange
quarks with roughly equal splitting. In Fig.~\ref{pic_mass}, the
masses of the $\sigma(600)$, $\kappa(800)$, $a_0(980)$ and
$f_0(980)$ are shown as functions of the Borel mass $M_B$. As we
see, the mass is very stable in a rather wide region of Borel mass
$M_B$.

%
\begin{table}
\caption{Masses of scalar nonet.}
\begin{center}
\begin{tabular}{c|c|c|c|c}
\hline Mass (MeV) & $\sigma(600)$ & $\kappa(800)$ & $a_0(980)$ &
$f_0(980)$
\\ \hline Experiments (PDG) & $400 \sim 1200$ & $841\pm30^{\Large +81}_{\Large -73}$ & $984.7\pm1.2$ &
$980\pm10$
\\ \hline QCD sum rule & $600 \pm 100$ & $800\pm100$ & $1000\pm100$ & $1000\pm100$
\\ \hline
\end{tabular}\label{table_mass}
\end{center}
\end{table}
%

%
%

The current $\eta_1$ has the antisymmetric flavor structure and
$\eta_2$ has the symmetric flavor structure. By using these currents
with different flavor structures, we arrive at similar QCD sum rule
results. This suggests that the tetraquarks of different flavor
structure may mix with each other, and the tetraquark states can
contain diquark and antidiquark having the mixing of the symmetric
flavor $\mathbf{6_f} \otimes \mathbf{\bar 6_f}$ and the
antisymmetric flavor $\mathbf{\bar 3_f} \otimes \mathbf{3_f}$, just
like they can have a mixing of different color, spin and orbital
symmetries. This is very much different from the ground baryon
states, where the different flavor representations $\mathbf{8}$ and
$\mathbf{10}$ correspond to different spins $\mathbf{1 / 2}$ and
$\mathbf{3 / 2}$, which induces a mass splitting between
$\Delta(1232)$ and $N(939)$.

%
\section{Finite Decay Width}\label{sec_decay}
%

The scalar mesons have large decay widthes, and it is important to
consider their effect. In this section, we use a Gaussian
distribution for the phenomenal spectral density, instead of
$\delta$-function,
%
\begin{eqnarray}
\rho^{FDW}(\sqrt{s}) d \sqrt{s} & \equiv & \sum_n \langle 0| \eta| n
\rangle\langle n|{\eta^\dagger}|0 \rangle {1 \over \sqrt{2 \pi}
\sigma } \exp \big ( - { (\sqrt{s} - M_n)^2 \over 2 * \sigma_n^2}
\big ) d\sqrt{s} \nonumber\\ &=&  {f^2_X \over \sqrt{2 \pi} \sigma }
\exp \big ( - { (\sqrt{s} - M_X)^2 \over 2 * \sigma_X^2} \big )
d\sqrt{s} + \rm{higher\,\,states} , \label{eq_decay}
\end{eqnarray}
%
where as usual the lowest state denoted by $X$ is isolated from the
rest of higher states. The Gaussian width $\sigma_X$ is related to
the Breit-Wigner decay width $\Gamma$ by $\sigma_X = \Gamma / 2.4$.

Again we assume the continuum contribution can be approximated by
the spectral density of OPE above a threshold value $s_0$, and we
arrive at the sum rule equation for state having a finite decay
width
%
\begin{equation}
\Pi^{FDW}(M_B^2) \equiv \int_{-\infty}^{+\infty} e^{-s/M_B^2} {1
\over \sqrt{2 \pi} \sigma } \exp \big ( - { (\sqrt{s} - M_X)^2 \over
2 \sigma_X^2} \big ) d\sqrt{s} = \int^{s_0}_0 e^{-s/M_B^2}\rho(s)ds
\label{eq_FDW} \, .
\end{equation}
%
For a given $\Gamma$, the mass can be obtained by solving the
equation
%
\begin{equation}
{ \int_{-\infty}^{+\infty} e^{-s/M_B^2} s \exp \big ( - { (\sqrt{s}
- M_X)^2 \over 2 \sigma_X^2} \big ) d\sqrt{s} \over
\int_{-\infty}^{+\infty} e^{-s/M_B^2} \exp \big ( - { (\sqrt{s} -
M_X)^2 \over 2 \sigma_X^2} \big ) d\sqrt{s} } e = { \int^{s_0}_0
e^{-s/M_B^2} s \rho(s) ds \over \int^{s_0}_0 e^{-s/M_B^2}\rho(s) ds
\label{eq_mass_FDW}}\, .
\end{equation}
%

\begin{figure}[hbt]
\begin{center}
\scalebox{0.8}{\includegraphics{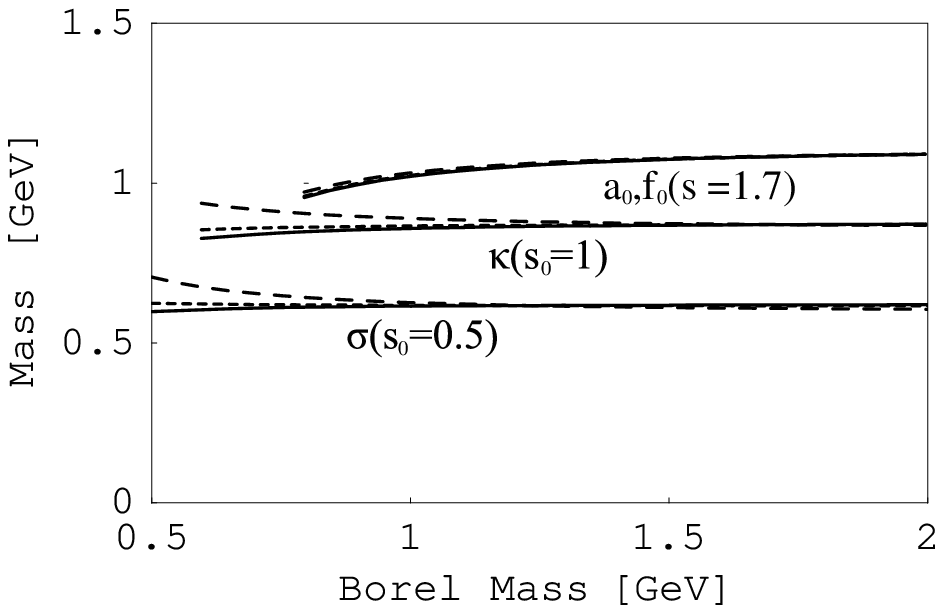}}
\scalebox{0.8}{\includegraphics{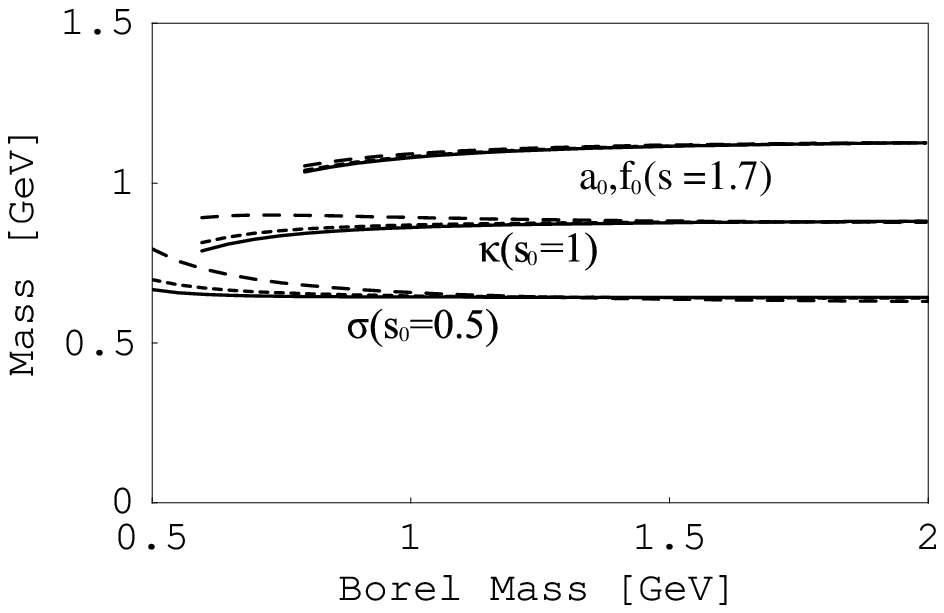}} \caption{Masses of
the $\sigma$, $\kappa$, $a_0$ and $f_0$ as tetraquark states
calculated by the mixed currents $\eta_1$ (left) and $\eta_2$
(right), as functions of the Borel mass $M_B$. For $\sigma$ and
$\kappa$, the solid, short-dashed and long-dashed curves are
obtained by setting $\Gamma = 0,~200$ and $400$ MeV respectively.
For $a_0$ and $f_0$, the solid, short-dashed and long-dashed curves
are obtained by setting $\Gamma = 0,~100$ and $200$ MeV
respectively.} \label{pic_decay}
\end{center}
\end{figure}

In Fig.~\ref{pic_decay}, the masses of the $\sigma(600)$,
$\kappa(800)$, $a_0(980)$ and $f_0(980)$ are shown as functions of
the Borel mass $M_B$, by setting $\Gamma = 0,~100,~200$ and $400$
MeV respectively. We find that after considering the finite decay
width by using the Gaussian distribution, the predicted masses do
not change significantly as far as the Borel mass is within a
reasonable range, where we can still reproduce the experimental
data. However, the question of finite decay width is very important,
and we do not consider that our attempt to use the Gaussian form is
the final. We need further investigations, which we would like to
put as a future important work.
%
%

%
\section{Conventional $\bar q q$ Mesons}\label{sec_meson}
%

For comparison, we have also performed the QCD sum rule analysis
using the $\bar q q$ current within the present framework. The QCD
sum rule analyses of conventional $\bar q q$ mesons have been
performed in
Ref.~\cite{Reinders:1981ww,Kisslinger:1997gs,Elias:1998bq,Du:2004ki}.
The sum rules using the current $j = \bar q_1 q_2$ are
%
\begin{eqnarray}
\nonumber f^2_{(q_1q_2)} e^{ -{m^2_{(q_1q_2)} \over M_B^2} } &=&
\int_0^{s_0} e^{-s/M_B^2} {3 \over 8 \pi^2} s \Big (1 + {17 \over
3}{\alpha_s \over \pi} \Big ) ds + {3 \over 2} \Big (m_1 \langle
\bar q_2 q_2 \rangle + m_2 \langle \bar q_1 q_1 \rangle \Big ) \\
&& + {1 \over 8 \pi} \langle {g^2 \over 4 \pi} G^2 \rangle - {1
\over 2 M_B^2} \Big ( m_1 \langle g \bar q_2 \sigma G q_2 \rangle +
m_2 \langle g \bar q_1 \sigma G q_1 \rangle \Big )
\\ \nonumber && - {16 \pi \over 3 M_B^2} {g_s \over 4 \pi }{\langle \bar q_1 q_1 \rangle
\langle \bar q_2 q_2 \rangle} - {16 \pi \over 27 M_B^2} {g_s \over 4
\pi } \Big ( {\langle \bar q_1 q_1 \rangle^2 + \langle \bar q_2 q_2
\rangle}^2 \Big ) \, .
\end{eqnarray}
%

The masses of $\sigma$ and $a_0$ are predicted to be around 1.2 GeV,
while the masses of $\kappa$ and $f_0$ are larger due to the
$strange$ quark content. Here again we have tested other values of
$M_B$ and $s_0$, and confirmed that the result shown is optimal.
These results are consistent with the previous
work~\cite{Reinders:1981ww,Kisslinger:1997gs,Elias:1998bq,Du:2004ki}.

%
\section{Summary}\label{sec_summary}
%

We have performed the QCD sum rule analysis with tetraquark
currents, and found the masses of scalar mesons in the region of 600
-- 1000 MeV with the ordering, $m_\sigma < m_\kappa < m_{f_0, a_0}$.
We have also used the conventional $\bar q q$ currents, and verified
their masses around 1.2 GeV. We have tested all possible independent
tetraquark currents as well as their linear combinations, and
considered the effect of finite decay width. Our conclusions are,
therefore, rather robust.

The scalar tetraquark currents can have either the antisymmetric
flavor or the symmetric flavor structures. We found that there are
five independent currents for each state. We investigated Borel mass
$M_B$ and threshold value $s_0$ dependences, which are quite stable.
The convergence of the OPE is also good, the positivity (of spectral
density) is maintained, and the pole contribution is sufficient
large. Therefore, we have achieved a QCD sum rule which is the best
reliable within the present calculation of OPE.

Our calculation supports a tetraquark structure for low-lying scalar
mesons. We find that the gluon condensate is quite large in the OPE
of the mixed currents, which is related to the question of the
origin of the mass generation of hadrons~\cite{Weinberg:1969hw}. We
obtain similar results by using the currents having both the
antisymmetric flavor structure and the symmetric flavor structure.
This suggests that the tetraquark can have a mixing of different
flavor symmetries, as well as different color, spin and orbital
symmetries. There is a mass splitting due to the different flavor,
color, spin and orbital structures. If this mass spitting is large
enough to be observed in experiments, the tetraquark spectrum would
become much more complicated; If the mass splitting is too small to
be observed in experiments, a broad decay width would be observed.
Such a tetraquark structure will open an alternative path toward the
understanding of exotic multiquark dynamics which one does not
experience in the conventional hadrons.

%
\section*{Acknowledgments}
%

The authors thank the M.~Oka, G.~Erkol, H.~J.~Lee and S.~H.~Lee for
useful discussions. H.X.C. is grateful to the Monkasho fellowship
for supporting his stay at Research Center for Nuclear Physics where
this work is done. A.H. is supported in part by the Grant for
Scientific Research ((C) No.19540297) from the Ministry of
Education, Culture, Science and Technology, Japan. S.L.Z. was
supported by the National Natural Science Foundation of China under
Grants 10421503 and 10625521, Ministry of Education of China,
FANEDD, Key Grant Project of Chinese Ministry of Education (NO
305001) and SRF for ROCS, SEM.

%
\appendix
%

%
\section{Relations between $(qq)(\bar q \bar q)$ and $(\bar q q)(\bar q q)$ Structures}\label{app_relation}
%

In this appendix, we study the relations between $(qq)(\bar q \bar
q)$ and $(\bar q q)(\bar q q)$ currents. We work under
$SU(3)_c\otimes~SU(3)_f\otimes~SO(1,3)_L$, where the quark field
$q^A_{a \mu}$ has the color index $a$, flavor index $A$ and Lorentz
index $\mu$. First, we consider the color and flavor structures. The
interchange of both color and flavor does not need to be
antisymmetric, due to the extra orbital and spin degrees of freedom.
Therefore we can not use the Pauli principle such as $q^A_a q^B_b =
- q^B_b q^A_a$ within the color and flavor spaces. Altogether there
are four types of diquark ($qq$) and four types of quark-antiquark
($\bar q q$). They are shown in Table~\ref{table_cf}, where the sum
over repeated indices ($a, b, \cdots$ for color indices, $A, B,
\cdots$ for flavor indices) is taken.
%
\begin{table}[h]
\begin{center}
\caption{Color and flavor structures of $qq$ and $\bar q q$}
\begin{tabular}{c|c|c|c|c}
\hline \hline & & & & \\ (Color, Flavor) & ($\mathbf{\bar 3_c}$,
$\mathbf{\bar 3_f}$) & ($\mathbf{\bar 3_c}$, $\mathbf{6_f}$) &
($\mathbf{6_c}$, $\mathbf{\bar 3_f}$) & ($\mathbf{6_c}$,
$\mathbf{6_f}$)
\\ \hline & & & & \\ Diquark ($qq$) & $\epsilon^{abc}\epsilon_{ABC} (q^A_a q^B_b)$ & $\epsilon^{abc} (q^A_a q^B_b + q^B_a q^A_b)$
& $\epsilon_{ABC} (q^A_a q^B_b + q^A_b q^B_a)$ & $(q^A_a q^B_b +
q^B_a q^A_b) + (a \leftrightarrow b)$
\\ \hline
\hline & & & & \\ (Color, Flavor) & ($\mathbf{1_c}$, $\mathbf{1_f}$)
& ($\mathbf{1_c}$, $\mathbf{8_f}$) & ($\mathbf{8_c}$,
$\mathbf{1_f}$) & ($\mathbf{8_c}$, $\mathbf{8_f}$)
\\ \hline & & & & \\ Quark-antiquark ($\bar q q$) & $(\bar q^A_a q^A_a)$ & $\lambda^N_{AB} (\bar q^A_a q^B_a)$ &
$\lambda_n^{ab} (\bar q^A_a q^A_b)$ & $\lambda^N_{AB} \lambda_n^{ab}
(\bar q^A_a q^B_b)$
\\ \hline \hline
\end{tabular}\label{table_cf}
\end{center}
\end{table}
%

To construct a tetraquark by using $(qq)(\bar q \bar q)$, the color
is either $(\mathbf{3} \otimes \mathbf{3}) \otimes (\mathbf{\bar 3}
\otimes \mathbf{\bar 3}) \rightarrow \mathbf{\bar 3} \otimes
\mathbf{3} \rightarrow \mathbf{1}$ or $(\mathbf{3} \otimes
\mathbf{3}) \otimes (\mathbf{\bar 3} \otimes \mathbf{\bar 3})
\rightarrow \mathbf{6} \otimes \mathbf{\bar 6} \rightarrow
\mathbf{1}$; the flavor is $(\mathbf{3} \otimes \mathbf{3}) \otimes
(\mathbf{\bar 3} \otimes \mathbf{\bar 3}) = (\mathbf{\bar 3} \oplus
\mathbf{6}) \otimes ( \mathbf{3} \oplus \mathbf{\bar 6}) =
\mathbf{1} \oplus \mathbf{8} \oplus \mathbf{8} \oplus \mathbf{10}
\oplus \mathbf{8} \oplus \mathbf{10} \oplus \mathbf{1} \oplus
\mathbf{8} \oplus \mathbf{27}$; To construct a tetraquark by using
$(\bar q q)(\bar q q)$, the color is either $(\mathbf{\bar 3}
\otimes \mathbf{3}) \otimes (\mathbf{\bar 3} \otimes \mathbf{3})
\rightarrow \mathbf{1} \otimes \mathbf{1} \rightarrow \mathbf{1}$ or
$(\mathbf{\bar 3} \otimes \mathbf{3}) \otimes (\mathbf{\bar 3}
\otimes \mathbf{3}) \rightarrow \mathbf{8} \otimes \mathbf{\bar 8}
\rightarrow \mathbf{1}$, with the same flavor structure as before.
In Table~\ref{table_cf_tetra}, we show all possible color and flavor
structures of tetraquark currents $T^{F_1(F_2)}_C$. Here $F_1$
denotes the flavor representation of tetraquark; $F_2$ and $C$ show
the intermediate flavor and color representations of either diquark
(antidiquark) or quark-antiquark. $S^{ABCD}$ is the totally
symmetric matrix. Because we want to make a scalar tetraquark state,
the diquark and antidiquark fields should have the same color, spin
and orbital symmetries. Therefore, they must have the same flavor
symmetry, which is either symmetric ($\mathbf{6_f} \otimes
\mathbf{\bar 6_f}$) or antisymmetric ($\mathbf{\bar 3_f} \otimes
\mathbf{3_f}$).
%
\begin{table}[h]
\begin{center}
\caption{Color and flavor structures of tetraquark currents}
\begin{tabular}{c|c|c} \hline \hline &
& \\ $(qq)(\bar q \bar q)$ & $(\mathbf{3} \otimes \mathbf{3})
\otimes (\mathbf{\bar 3} \otimes \mathbf{\bar 3}) \rightarrow
\mathbf{\bar 3} \otimes \mathbf{3} \rightarrow \mathbf{1_c}$ &
$(\mathbf{3} \otimes \mathbf{3}) \otimes (\mathbf{\bar 3} \otimes
\mathbf{\bar 3}) \rightarrow \mathbf{6} \otimes \mathbf{\bar 6}
\rightarrow \mathbf{1_c}$
\\ \hline & & \\ $(\mathbf{3} \otimes \mathbf{3}) \otimes (\mathbf{\bar
3} \otimes \mathbf{\bar 3})$ &
$\epsilon^{abe}\epsilon^{cde}\epsilon_{ABE}\epsilon_{CDE} (q^A_a
q^B_b)(\bar q^C_c \bar q^D_d) \equiv
T^{\mathbf{1}(\mathbf{3})}_{\mathbf{3}}$ &
$\epsilon_{ABE}\epsilon_{CDE} (q^A_a q^B_b + q^A_b q^B_a)(\bar q^C_a
\bar q^D_b + \bar q^C_b \bar q^D_a)$
\\ $\rightarrow \mathbf{\bar 3} \otimes
\mathbf{3} \rightarrow \mathbf{1_f}$ & & $ = 2
\epsilon_{ABE}\epsilon_{CDE} (q^A_a q^B_b)(\bar q^C_a \bar q^D_b +
\bar q^C_b \bar q^D_a) \equiv 2
T^{\mathbf{1}(\mathbf{3})}_{\mathbf{6}}$
\\ \hline & & \\ $\rightarrow \mathbf{\bar 3} \otimes
\mathbf{3} \rightarrow \mathbf{8_f}$ &
$\epsilon^{abe}\epsilon^{cde}\lambda_N^{EF}\epsilon_{ABE}\epsilon_{CDF}
(q^A_a q^B_b)(\bar q^C_c \bar q^D_d) \equiv
T^{\mathbf{8}(\mathbf{3})}_{\mathbf{3}}$ &
$\lambda_N^{EF}\epsilon_{ABE} \epsilon_{CDF} (q^A_a q^B_b)(\bar
q^C_a \bar q^D_b + \bar q^C_b \bar q^D_a) \equiv
T^{\mathbf{8}(\mathbf{3})}_{\mathbf{6}}$
\\ \hline & & \\ $\rightarrow \mathbf{\bar 3} \otimes
\mathbf{\bar 6} \rightarrow \mathbf{8_f}$ &
$\epsilon^{abe}\epsilon^{cde}\lambda_N^{DF}\epsilon_{ABE}\epsilon_{CEF}
(q^A_a q^B_b)(\bar q^C_c \bar q^D_d) \equiv
T^{\mathbf{8}(\mathbf{3},\mathbf{6})}_{\mathbf{3}}$ &
$\lambda_N^{DF}\epsilon_{ABE} \epsilon_{CEF} (q^A_a q^B_b)(\bar
q^C_a \bar q^D_b + \bar q^C_b \bar q^D_a) \equiv
T^{\mathbf{8}(\mathbf{3},\mathbf{6})}_{\mathbf{6}}$
\\ \hline & & \\ $\rightarrow \mathbf{\bar 3} \otimes
\mathbf{\bar 6} \rightarrow \mathbf{10_f}$ &
$\epsilon^{abe}\epsilon^{cde} S^{CDE}\epsilon_{ABE} (q^A_a
q^B_b)(\bar q^C_c \bar q^D_d) \equiv
T^{\mathbf{10}(\mathbf{3},\mathbf{6})}_{\mathbf{3}}$ &
$S^{CDE}\epsilon_{ABE} (q^A_a q^B_b)(\bar q^C_a \bar q^D_b + \bar
q^C_b \bar q^D_a) \equiv
T^{\mathbf{10}(\mathbf{3},\mathbf{6})}_{\mathbf{6}}$
\\ \hline & & \\ $\rightarrow \mathbf{6} \otimes
\mathbf{3} \rightarrow \mathbf{8_f}$ &
$\epsilon^{abe}\epsilon^{cde}\lambda_N^{BF}\epsilon_{AEF}\epsilon_{CDE}
(q^A_a q^B_b)(\bar q^C_c \bar q^D_d) \equiv
T^{\mathbf{8}(\mathbf{6},\mathbf{3})}_{\mathbf{3}}$ &
$\lambda_N^{BF}\epsilon_{AEF} \epsilon_{CDE} (q^A_a q^B_b)(\bar
q^C_a \bar q^D_b + \bar q^C_b \bar q^D_a) \equiv
T^{\mathbf{8}(\mathbf{6},\mathbf{3})}_{\mathbf{6}}$
\\ \hline & & \\ $\rightarrow \mathbf{6} \otimes
\mathbf{3} \rightarrow \mathbf{10_f}$ &
$\epsilon^{abe}\epsilon^{cde}S^{ABE}\epsilon_{CDE} (q^A_a
q^B_b)(\bar q^C_c \bar q^D_d) \equiv
T^{\mathbf{10}(\mathbf{6},\mathbf{3})}_{\mathbf{3}}$ &
$S^{ABE}\epsilon_{CDE} (q^A_a q^B_b)(\bar q^C_a \bar q^D_b + \bar
q^C_b \bar q^D_a) \equiv
T^{\mathbf{10}(\mathbf{6},\mathbf{3})}_{\mathbf{6}}$
\\ \hline & & \\ $\rightarrow \mathbf{6} \otimes
\mathbf{\bar 6} \rightarrow \mathbf{1_f}$ &
$\epsilon^{abe}\epsilon^{cde}(q^A_a q^B_b + q^B_a q^A_b)(\bar q^A_c
\bar q^B_d + \bar q^B_c \bar q^A_d)$ & $(q^A_a q^B_b + q^B_a q^A_b
)(\bar q^A_a \bar q^B_b + \bar q^B_a \bar q^A_b + (a \leftrightarrow
b ))$
\\  & $ = 2
\epsilon^{abe}\epsilon^{cde}(q^A_a q^B_b)(\bar q^A_c \bar q^B_d +
\bar q^B_c \bar q^A_d) \equiv 2
T^{\mathbf{1}(\mathbf{6})}_{\mathbf{3}} $ & $ = 2 (q^A_a q^B_b)(\bar
q^A_a \bar q^B_b + \bar q^B_a \bar q^A_b + ( a \leftrightarrow b))
\equiv 2 T^{\mathbf{1}(\mathbf{6})}_{\mathbf{6}}$
\\ \hline & & \\ $\rightarrow \mathbf{6} \otimes
\mathbf{\bar 6} \rightarrow \mathbf{8_f}$ & $\lambda^N_{BC}
\epsilon^{abe}\epsilon^{cde}(q^A_a q^B_b + q^B_a q^A_b)(\bar q^A_c
\bar q^C_d + \bar q^C_c \bar q^A_d) \equiv
T^{\mathbf{8}(\mathbf{6})}_{\mathbf{3}}$ & $\lambda^N_{BC} (q^A_a
q^B_b + q^B_a q^A_b )(\bar q^A_a \bar q^C_b + \bar q^C_a \bar q^A_b
+ ( a \leftrightarrow b)) \equiv
T^{\mathbf{8}(\mathbf{6})}_{\mathbf{6}}$
\\ \hline & & \\ $\rightarrow \mathbf{6} \otimes
\mathbf{\bar 6} \rightarrow \mathbf{27_f}$ & $S_{ABCD}
\epsilon^{abe}\epsilon^{cde}(q^A_a q^B_b)(\bar q^C_c \bar q^D_d)
\equiv T^{\mathbf{27}(\mathbf{6})}_{\mathbf{3}}$ & $S_{ABCD} (q^A_a
q^B_b)(\bar q^C_a \bar q^D_b + \bar q^C_b \bar q^D_a) \equiv
T^{\mathbf{27}(\mathbf{6})}_{\mathbf{6}}$
\\ \hline \hline &
& \\ $(\bar q q)(\bar q q)$ & $(\mathbf{\bar 3} \otimes \mathbf{3})
\otimes (\mathbf{\bar 3} \otimes \mathbf{3}) \rightarrow \mathbf{1}
\otimes \mathbf{1} \rightarrow \mathbf{1_c}$ & $(\mathbf{\bar 3}
\otimes \mathbf{3}) \otimes (\mathbf{\bar 3} \otimes \mathbf{3})
\rightarrow \mathbf{8} \otimes \mathbf{8} \rightarrow \mathbf{1_c}$
\\ \hline & & \\ $(\mathbf{\bar 3} \otimes \mathbf{3}) \otimes (\mathbf{\bar 3} \otimes
\mathbf{3})$ & $(\bar q^A_a q^A_a)(\bar q^B_b q^B_b) \equiv
T^{\mathbf{1}(\mathbf{1})}_{\mathbf{1}}$ & $(\bar q^A_a
\lambda_n^{ab} q^A_b)(\bar q^B_c \lambda_n^{cd} q^B_d) \equiv
T^{\mathbf{1}(\mathbf{1})}_{\mathbf{8}}$
\\ $\rightarrow \mathbf{1} \otimes
\mathbf{1} \rightarrow \mathbf{1_f}$ & &
\\ \hline & & \\ $\rightarrow \mathbf{1} \otimes
\mathbf{8} \rightarrow \mathbf{8_f}$ & $\lambda^N_{BC} (\bar q^A_a
q^A_a)(\bar q^B_b q^C_b) \equiv
T^{\mathbf{8}(\mathbf{1},~\mathbf{8})}_{\mathbf{1}}$ &
$\lambda^N_{BC} (\bar q^A_a \lambda_n^{ab} q^A_b)(\bar q^B_c
\lambda_n^{cd} q^C_d) \equiv
T^{\mathbf{8}(\mathbf{1},~\mathbf{8})}_{\mathbf{8}}$
\\ \hline & & \\ $\rightarrow \mathbf{8} \otimes
\mathbf{1} \rightarrow \mathbf{8_f}$ & $\lambda^N_{BC} (\bar q^B_a
q^C_a)(\bar q^A_b q^A_b) \equiv
T^{\mathbf{8}(\mathbf{8},~\mathbf{1})}_{\mathbf{1}}$ &
$\lambda^N_{BC} (\bar q^B_a \lambda_n^{ab} q^C_b)(\bar q^A_c
\lambda_n^{cd} q^A_d) \equiv
T^{\mathbf{8}(\mathbf{8},~\mathbf{1})}_{\mathbf{8}}$
\\ \hline & & \\ $\rightarrow \mathbf{8} \otimes
\mathbf{8} \rightarrow \mathbf{1_f}$ & $(\bar q^A_a \lambda^N_{AB}
q^B_a)(\bar q^C_b \lambda^N_{CD} q^D_b) \equiv
T^{\mathbf{1}(\mathbf{8})}_{\mathbf{1}}$ & $(\bar q^A_a
\lambda_n^{ab} \lambda^N_{AB} q^B_b)(\bar q^C_c \lambda_n^{cd}
\lambda^N_{CD} q^D_d) \equiv
T^{\mathbf{1}(\mathbf{8})}_{\mathbf{8}}$
\\ \hline & & \\ $\rightarrow \mathbf{8} \otimes
\mathbf{8} \rightarrow \mathbf{8_f}$ & $\lambda_N^{FE}
\epsilon_{ACE} \epsilon_{BDF} (\bar q^A_a q^B_a)(\bar q^C_b q^D_b)
\equiv T^{\mathbf{8}(\mathbf{8})}_{\mathbf{1}}$ & $\lambda_N^{FE}
\epsilon_{ACE} \epsilon_{BDF} (\bar q^A_a \lambda_n^{ab} q^B_b)(\bar
q^C_c \lambda_n^{cd} q^D_d) \equiv
T^{\mathbf{8}(\mathbf{8})}_{\mathbf{8}}$
\\ \hline & & \\ $\rightarrow \mathbf{8} \otimes
\mathbf{8} \rightarrow \mathbf{8_f^\prime}$  & $\lambda_N^{BF}
\epsilon_{ACE} \epsilon_{DEF} (\bar q^A_a q^B_a)(\bar q^C_b q^D_b)
\equiv T^{\mathbf{8^\prime}(\mathbf{8})}_{\mathbf{1}}$ &
$\lambda_N^{BF} \epsilon_{ACE} \epsilon_{DEF} (\bar q^A_a
\lambda_n^{ab} q^B_b)(\bar q^C_c \lambda_n^{cd} q^D_d) \equiv
T^{\mathbf{8^\prime}(\mathbf{8})}_{\mathbf{8}}$
\\ \hline & & \\ $\rightarrow \mathbf{8} \otimes
\mathbf{8} \rightarrow \mathbf{10_f}$ & $\epsilon_{ACE} S_{BDE}
(\bar q^A_a q^B_a)(\bar q^C_b q^D_b) \equiv
T^{\mathbf{10}(\mathbf{8})}_{\mathbf{1}}$ & $\epsilon_{ACE} S_{BDE}
(\bar q^A_a \lambda_n^{ab} q^B_b)(\bar q^C_c \lambda_n^{cd} q^D_d)
\equiv T^{\mathbf{10}(\mathbf{8})}_{\mathbf{8}}$
\\ \hline & & \\ $\rightarrow \mathbf{8} \otimes
\mathbf{8} \rightarrow \mathbf{10_f^\prime}$ & $\epsilon_{BDE}
S_{ACE} (\bar q^A_a q^B_a)(\bar q^C_b q^D_b) \equiv
T^{\mathbf{10^\prime}(\mathbf{8})}_{\mathbf{1}}$ & $\epsilon_{BDE}
S_{ACE} (\bar q^A_a \lambda_n^{ab} q^B_b)(\bar q^C_c \lambda_n^{cd}
q^D_d) \equiv T^{\mathbf{10^\prime}(\mathbf{8})}_{\mathbf{8}}$
\\ \hline & & \\ $\rightarrow \mathbf{8} \otimes
\mathbf{8} \rightarrow \mathbf{27_f}$ & $S_{ABCD} (\bar q^A_a
q^B_a)(\bar q^C_b q^D_b) \equiv
T^{\mathbf{27}(\mathbf{8})}_{\mathbf{1}}$ & $S_{ABCD} (\bar q^A_a
\lambda_n^{ab} q^B_b)(\bar q^C_c \lambda_n^{cd} q^D_d) \equiv
T^{\mathbf{27}(\mathbf{8})}_{\mathbf{8}}$
\\ \hline \hline
\end{tabular}\label{table_cf_tetra}
\end{center}
\end{table}
%

If the orbital and spin structure between the two quarks (two
antiquarks) are symmetric, then the color-flavor structure of
diquark (antidiquark) should be anti-symmetric, which means $q^A_a
q^B_b = - q^B_b q^A_a$ ($\bar q^A_a \bar q^B_b = - \bar q^B_b \bar
q^A_a$). In this case, we can verify
%
\begin{equation}\label{eq_antisymmetric}
T^{\mathbf{1}(\mathbf{3})}_\mathbf{3} =
T^{\mathbf{8}(\mathbf{3})}_\mathbf{3} =
T^{\mathbf{8}(\mathbf{3},\mathbf{6})}_\mathbf{3} =
T^{\mathbf{10}(\mathbf{3},\mathbf{6})}_\mathbf{3} =
T^{\mathbf{8}(\mathbf{6},\mathbf{3})}_\mathbf{6} =
T^{\mathbf{10}(\mathbf{6},\mathbf{3})}_\mathbf{6} =
T^{\mathbf{1}(\mathbf{6})}_\mathbf{6} =
T^{\mathbf{8}(\mathbf{6})}_\mathbf{6} =
T^{\mathbf{27}(\mathbf{6})}_\mathbf{6} = 0 \, ,
\end{equation}
%
If the orbital and spin structure between two quarks (two
antiquarks) are anti-symmetric, then the color-flavor structure of
diquark (antidiquark) should be symmetric, which means $q^A_a q^B_b
= q^B_b q^A_a$ ($\bar q^A_a \bar q^B_b = \bar q^B_b \bar q^A_a$).
Then we can verify
%
\begin{equation}\label{eq_symmetric}
T^{\mathbf{1}(\mathbf{3})}_\mathbf{6} =
T^{\mathbf{8}(\mathbf{3})}_\mathbf{6} =
T^{\mathbf{8}(\mathbf{3},\mathbf{6})}_\mathbf{6} =
T^{\mathbf{10}(\mathbf{3},\mathbf{6})}_\mathbf{6} =
T^{\mathbf{8}(\mathbf{6},\mathbf{3})}_\mathbf{3} =
T^{\mathbf{10}(\mathbf{6},\mathbf{3})}_\mathbf{3} =
T^{\mathbf{1}(\mathbf{6})}_\mathbf{3} =
T^{\mathbf{8}(\mathbf{6})}_\mathbf{3} =
T^{\mathbf{27}(\mathbf{6})}_\mathbf{3} = 0 \, .
\end{equation}
%

Now let us discuss the Fierz rearrangement in order to relate $(q
q)(\bar q \bar q)$ and $(\bar q q)(\bar q q)$ structures. First we
perform it in the color and flavor spaces. To do this, it is
convenient to consider the interchange of color indices:
%
\begin{eqnarray}\label{eq_color}\nonumber
&&(q^A_a q^B_b \bar q^C_a \bar q^D_b) = {1 \over 3} (q^A_a q^B_b
\bar q^C_b \bar q^D_a) + {1 \over 2} \lambda_n^{ab}\lambda_n^{cd}
(q^A_a q^B_c \bar q^C_d \bar q^D_b) \, ,
\\ && \lambda_n^{ab}\lambda_n^{cd}(q^A_a q^B_c \bar q^C_b \bar q^D_d) = {16 \over 9} (q^A_a
q^B_b \bar q^C_b \bar q^D_a) - {1 \over 3}
\lambda_n^{ab}\lambda_n^{cd} (q^A_a q^B_c \bar q^C_d \bar q^D_b) \,
.
\end{eqnarray}
%
We can obtain the same result for flavor structure.

Let us take $T^{\mathbf{1}(\mathbf{3})}_{\mathbf{3}}$ as an example,
and perform the simultaneous interchange of both color and flavor
indices
%
\begin{eqnarray}\nonumber
T^{\mathbf{1}(\mathbf{3})}_{\mathbf{3}} &=&
\epsilon^{abe}\epsilon^{cde}\epsilon_{ABE}\epsilon_{CDE} (q^A_a
q^B_b)(\bar q^C_c \bar q^D_d) \\ \nonumber &=& (q^A_a q^B_b)(\bar
q^A_a \bar q^B_b) - (q^A_a q^B_b)(\bar q^A_b \bar q^B_a) - (q^A_a
q^B_b)(\bar q^B_a \bar q^A_b) + (q^A_a q^B_b)(\bar q^B_b \bar q^A_a)
\\ \nonumber &=& (q^A_a q^B_b)(\bar
q^A_a \bar q^B_b) - \Big ({1 \over 3}(q^A_a q^B_b)(\bar q^A_a \bar
q^B_b) + {1 \over 2} \lambda_n^{ab}\lambda_n^{cd} (q^A_a q^B_c)(\bar
q^A_b \bar q^B_d) \Big ) \\ \nonumber && - (q^A_a q^B_b)(\bar q^B_a
\bar q^A_b) + \Big ( {1 \over 3}(q^A_a q^B_b)(\bar q^B_a \bar q^A_b)
+ {1 \over 2} \lambda_n^{ab}\lambda_n^{cd} (q^A_a q^B_c)(\bar q^B_b
\bar q^A_d) \Big )
\\ \nonumber &=& {2 \over 3}(q^A_a q^B_b)(\bar q^A_a \bar
q^B_b) - {1 \over 2} \lambda_n^{ab}\lambda_n^{cd} (q^A_a q^B_c)(\bar
q^A_b \bar q^B_d) -  {2 \over 3} \Big ( {1 \over 3}(q^A_a
q^B_b)(\bar q^A_a \bar q^B_b) + {1 \over 2}
\lambda^N_{AB}\lambda^N_{CD} (q^A_a q^C_b)(\bar q^B_a \bar q^D_b)
\Big ) \\ \nonumber &&  + {1 \over 2} \Big ( {1 \over
3}\lambda_n^{ab}\lambda_n^{cd} (q^A_a q^B_c)(\bar q^A_b \bar q^B_d)
+ {1 \over 2} \lambda^N_{AB}\lambda^N_{CD}
\lambda_n^{ab}\lambda_n^{cd} (q^A_a q^C_c)(\bar q^B_b \bar q^D_d)
\Big )
\\ \nonumber &=& {4 \over 9}(q^A_a q^B_b)(\bar q^A_a \bar
q^B_b) - {1 \over 3} \lambda_n^{ab}\lambda_n^{cd} (q^A_a q^B_c)(\bar
q^A_b \bar q^B_d) -  {1 \over 3} \lambda^N_{AB}\lambda^N_{CD} (q^A_a
q^C_b)(\bar q^B_a \bar q^D_b) \\ \nonumber &&  + {1 \over 4}
\lambda^N_{AB}\lambda^N_{CD} \lambda_n^{ab}\lambda_n^{cd} (q^A_a
q^C_c)(\bar q^B_b \bar q^D_d) \, .
\end{eqnarray}
Because we only consider the color and flavor structures, by
changing the ordering of the second quark and third quark, we arrive
at the result:
\begin{eqnarray}\label{ex_cf}\nonumber
&\sim& {4 \over 9}(\bar q^A_a q^A_a)(\bar q^B_b q^B_b) - {1 \over 3}
\lambda_n^{ab}\lambda_n^{cd} (\bar q^A_b q^A_a)(\bar q^B_d q^B_c) -
{1 \over 3}
\lambda^N_{AB}\lambda^N_{CD} (\bar q^B_a q^A_a)( \bar q^D_b q^C_b) \\
\nonumber &&  + {1 \over 4} \lambda^N_{AB}\lambda^N_{CD}
\lambda_n^{ab}\lambda_n^{cd} (\bar q^B_b q^A_a)(\bar q^D_d q^C_c) \,
.
\\ &=& {4 \over 9} T^{\mathbf{1}(\mathbf{1})}_{\mathbf{1}} - {1 \over 3}
T^{\mathbf{1}(\mathbf{1})}_{\mathbf{8}} - {1 \over 3}
T^{\mathbf{1}(\mathbf{8})}_{\mathbf{1}} + {1 \over 4}
T^{\mathbf{1}(\mathbf{8})}_{\mathbf{8}} \, .
\end{eqnarray}
%

Next we perform the Fierz rearrangement in the Lorentz indices. The
formulae is:
%
\begin{equation}
(\mathbf{1})_{\alpha\beta} (\mathbf{1})_{\gamma\delta} = {1 \over 4}
(\mathbf{1})_{\alpha\delta} (\mathbf{1})_{\gamma\beta} + {1 \over 4}
(\mathbf{\gamma_\mu})_{\alpha\delta}
(\mathbf{\gamma^\mu})_{\gamma\beta} + {1 \over 8}
(\mathbf{\sigma_{\mu\nu}})_{\alpha\delta}
(\mathbf{\sigma^{\mu\nu}})_{\gamma\beta} - {1 \over 4}
(\mathbf{\gamma_\mu \gamma_5})_{\alpha\delta} (\mathbf{\gamma^\mu
\gamma_5})_{\gamma\beta} + {1 \over 4}
(\mathbf{\gamma_5})_{\alpha\delta}
(\mathbf{\gamma_5})_{\gamma\beta}\, .
\end{equation}
%
By using this equation, we can obtain various relations such as
%
\begin{eqnarray} \nonumber
((q^A_a)^T C q^B_b)(\bar q^C_c C (\bar q^D_d)^T) &=& - {1 \over 4}
((q^A_a)^T C C (\bar q^D_d)^T) (\bar q^C_c q^B_b) - {1 \over 4}
((q^A_a)^T C \gamma_\mu
C (\bar q^D_d)^T ) (\bar q^C_c \gamma^\mu q^B_b) \\
\nonumber && - {1 \over 8} ((q^A_a)^T C \sigma_{\mu\nu} C (\bar
q^D_d)^T) (\bar q^C_c \sigma^{\mu\nu} q^B_b) + {1 \over 4}
((q^A_a)^T C \gamma_\mu \gamma_5 C (\bar q^D_d)^T) (\bar q^C_c
\gamma^\mu \gamma_5
q^B_b) \\
\nonumber && - {1 \over 4} ((q^A_a)^T C \gamma_5 C (\bar q^D_d)^T)
(\bar q^C_c \gamma_5 q^B_b) \\ \nonumber &=& - {1 \over 4} (\bar
q^D_d q^A_a) (\bar q^C_c q^B_b) + {1 \over 4} (\bar q^D_d \gamma_\mu
q^A_a) (\bar q^C_c \gamma^\mu q^B_b)  + {1 \over 8} (\bar q^D_d
\sigma_{\mu\nu} q^A_a) (\bar q^C_c \sigma^{\mu\nu} q^B_b)
\\
&& + {1 \over 4} (\bar q^D_d \gamma_\mu \gamma_5 q^A_a) (\bar q^C_c
\gamma^\mu \gamma_5 q^B_b) - {1 \over 4} (\bar q^D_d \gamma_5 q^A_a)
(\bar q^C_c \gamma_5 q^B_b)\, .
\end{eqnarray}
%

In order to label the Lorentz structure for a scalar tetraquark
field, we introduce $S$, $V$, $T$, $A$ and $P$ instead of $T$:
\begin{eqnarray} \nonumber &&
S \mbox{ for } (q^T C \gamma_5 q)(\bar q \gamma_5 C \bar q^T) \mbox{
and } (\bar q q)(\bar q q) \, , \\ \nonumber && V \mbox{ for } (q^T
C \gamma_\mu \gamma_5 q)(\bar q \gamma^\mu \gamma_5 C \bar
q^T) \mbox{ and } (\bar q \gamma_\mu q)(\bar q \gamma^\mu q) \, , \\
\nonumber && T \mbox{ for } (q^T C \sigma_{\mu\nu} q)(\bar q
\sigma^{\mu\nu} C \bar q^T) \mbox{ and } (\bar q
\sigma_{\mu\nu} q)(\bar q \sigma^{\mu\nu} q) \, , \\
\nonumber && A \mbox{ for } (q^T C \gamma_\mu q)(\bar q \gamma^\mu C
\bar q^T) \mbox{ and } (\bar q
\gamma_\mu \gamma_5 q)(\bar q \gamma^\mu \gamma_5 q) \, , \\
\nonumber && P \mbox{ for } (q^T C q)(\bar q C \bar q^T) \mbox{ and
} (\bar q \gamma_5 q)(\bar q \gamma_5 q) \, .
\end{eqnarray}
For example,
\begin{eqnarray} \nonumber &&
S^{\mathbf{27}(\mathbf{6})}_\mathbf{6} \equiv S_{ABCD} ( q^{AT}_a C
\gamma_5 q^B_b )(\bar q^C_a \gamma_5 C \bar q^{DT}_b + \bar q^C_b
\gamma_5 C \bar q^{DT}_a)
\, , \\
&& V^{\mathbf{27}(\mathbf{8})}_\mathbf{1} \equiv S_{ABCD} (\bar
q^A_a \gamma_\mu q^B_a )(\bar q^C_b \gamma^\mu q^D_b ) \, .
\end{eqnarray}

Diquarks belonging to $T$ and $A$ have a symmetric Lorentz structure
(see Eq.~\ref{eq_antisymmetric})
\begin{equation}
(C \gamma_\mu)_{\alpha\beta} =  (C \gamma_\mu)_{\beta\alpha} \, , (C
\sigma_{\mu\nu})_{\alpha\beta} = (C \sigma_{\mu\nu})_{\beta\alpha}
\, ,
\end{equation}
so they have an anti-symmetric color-flavor structure. Therefore,
currents having the symmetric color-flavor structure vanish, such as
%
\begin{equation}
A^{\mathbf{1}(\mathbf{3})}_\mathbf{3} = \epsilon^{abe}
\epsilon^{cde} \epsilon_{ABE} \epsilon_{CDE} ( (q^A_a)^T C
\gamma_\mu q^B_b) ( \bar q^C_c \gamma^\mu C (\bar q^D_d)^T ) = 0\, .
\end{equation}
%
Similarly, diquarks belonging to $S$, $V$ and $P$ have an
anti-symmetric Lorentz structure (see Eq.~\ref{eq_symmetric})
\begin{equation}
(C)_{\alpha\beta} = - (C)_{\beta\alpha} \, , (C \gamma_\mu
\gamma_5)_{\alpha\beta} = - (C \gamma_\mu \gamma_5)_{\beta\alpha} \,
, (C \gamma_5)_{\alpha\beta} = - (C \gamma_5)_{\beta\alpha} \, ,
\end{equation}
and so they have a symmetric color-flavor structure.

By now, we have known the flavor, color and Lorentz structures of
scalar tetraquark fields, for both $(qq)(\bar q \bar q)$ and $(\bar
q q)(\bar q q)$ structures, and are ready to derive some relations.

%
\subsection{Specifying the flavor structure}\label{app_flavor}
%

In order to establish the relations, we need to specify the flavor
quantum numbers of the tetraquark currents. As we are considering in
this work, let us choose the flavor octet states $(\mathbf{3}
\otimes \mathbf{3}) \otimes (\mathbf{\bar 3} \otimes \mathbf{\bar
3}) \rightarrow \mathbf{\bar 3} \otimes \mathbf{3} \rightarrow
\mathbf{8_f}$ for the illustration.

In this case, diquarks and antidiquarks have an anti-symmetric
flavor structure, and we can verify
%
\begin{equation}
S^{\mathbf{8}(\mathbf{3})}_\mathbf{6} =
V^{\mathbf{8}(\mathbf{3})}_\mathbf{6} =
T^{\mathbf{8}(\mathbf{3})}_\mathbf{3} =
A^{\mathbf{8}(\mathbf{3})}_\mathbf{3} =
P^{\mathbf{8}(\mathbf{3})}_\mathbf{6} = 0 \, .
\end{equation}
%
Therefore, there are five types of $(qq)(\bar q \bar q)$ fields
which are non-zero and independent:
%
\begin{equation} \nonumber
S^{\mathbf{8}(\mathbf{3})}_\mathbf{3} \, ,
V^{\mathbf{8}(\mathbf{3})}_\mathbf{3} \, ,
T^{\mathbf{8}(\mathbf{3})}_\mathbf{6} \, ,
A^{\mathbf{8}(\mathbf{3})}_\mathbf{6} \, ,
P^{\mathbf{8}(\mathbf{3})}_\mathbf{3} \, ,
\end{equation}
%
while all ten types remain for the $(\bar q q)(\bar q q)$ fields:
%
\begin{equation} \nonumber
S^{\mathbf{8}(\mathbf{8})}_\mathbf{1} \, ,
V^{\mathbf{8}(\mathbf{8})}_\mathbf{1} \, ,
T^{\mathbf{8}(\mathbf{8})}_\mathbf{1} \, ,
A^{\mathbf{8}(\mathbf{8})}_\mathbf{1} \, ,
P^{\mathbf{8}(\mathbf{8})}_\mathbf{1} \, ,
S^{\mathbf{8}(\mathbf{8})}_\mathbf{8} \, ,
V^{\mathbf{8}(\mathbf{8})}_\mathbf{8} \, ,
T^{\mathbf{8}(\mathbf{8})}_\mathbf{8} \, ,
A^{\mathbf{8}(\mathbf{8})}_\mathbf{8} \, ,
P^{\mathbf{8}(\mathbf{8})}_\mathbf{8} \, ,
\end{equation}
%
Among these ten $(\bar q q)(\bar q q)$ fields, only five are
independent. We can derive the following five equation by applying
the Fierz transformation for the $(\bar q q)(\bar q q)$ fields:
%
\begin{eqnarray}\label{eq_flavor_qantiq}\nonumber
S^{\mathbf{8}(\mathbf{8})}_\mathbf{8} &=&  -\frac{1}{6}
S^{\mathbf{8}(\mathbf{8})}_\mathbf{1} + \frac{1}{2}
V^{\mathbf{8}(\mathbf{8})}_\mathbf{1} + \frac{1}{4}
T^{\mathbf{8}(\mathbf{8})}_\mathbf{1} - \frac{1}{2}
A^{\mathbf{8}(\mathbf{8})}_\mathbf{1} - \frac{1}{2}
P^{\mathbf{8}(\mathbf{8})}_\mathbf{1}\, ,
\\ \nonumber V^{\mathbf{8}(\mathbf{8})}_\mathbf{8} &=& 2 S^{\mathbf{8}(\mathbf{8})}_\mathbf{1}
- \frac{5}{3} V^{\mathbf{8}(\mathbf{8})}_\mathbf{1} -
A^{\mathbf{8}(\mathbf{8})}_\mathbf{1} - 2
P^{\mathbf{8}(\mathbf{8})}_\mathbf{1}\, ,
\\ T^{\mathbf{8}(\mathbf{8})}_\mathbf{8} &=& 6 S^{\mathbf{8}(\mathbf{8})}_\mathbf{1}
- \frac{5}{3} T^{\mathbf{8}(\mathbf{8})}_\mathbf{1} + 6
P^{\mathbf{8}(\mathbf{8})}_\mathbf{1}\, ,
\\ \nonumber A^{\mathbf{8}(\mathbf{8})}_\mathbf{8} &=& - 2 S^{\mathbf{8}(\mathbf{8})}_\mathbf{1}
- V^{\mathbf{8}(\mathbf{8})}_\mathbf{1} - \frac{5}{3}
A^{\mathbf{8}(\mathbf{8})}_\mathbf{1} + 2
P^{\mathbf{8}(\mathbf{8})}_\mathbf{1}\, ,
\\ \nonumber P^{\mathbf{8}(\mathbf{8})}_\mathbf{8} &=&
\frac{1}{2} S^{\mathbf{8}(\mathbf{8})}_\mathbf{1} - \frac{1}{2}
V^{\mathbf{8}(\mathbf{8})}_\mathbf{1} + \frac{1}{4}
T^{\mathbf{8}(\mathbf{8})}_\mathbf{1} + \frac{1}{2}
A^{\mathbf{8}(\mathbf{8})}_\mathbf{1} - \frac{1}{6}
P^{\mathbf{8}(\mathbf{8})}_\mathbf{1}\, .
\end{eqnarray}
%

Employing the five currents on the left hand sides of
Eqs.~(\ref{eq_flavor_qantiq}) as independent ones, and applying the
Fierz transformation, we can establish the following relations among
the five $(qq)(\bar q \bar q)$ and five $(\bar q q)(\bar q q)$
structures:
%
\begin{eqnarray}
\nonumber S^{\mathbf{8}(\mathbf{3})}_\mathbf{3} &=& -\frac{1}{2}
S^{\mathbf{8}(\mathbf{8})}_\mathbf{1} - \frac{1}{2}
V^{\mathbf{8}(\mathbf{8})}_\mathbf{1} + \frac{1}{4}
T^{\mathbf{8}(\mathbf{8})}_\mathbf{1} - \frac{1}{2}
A^{\mathbf{8}(\mathbf{8})}_\mathbf{1}
- \frac{1}{2} P^{\mathbf{8}(\mathbf{8})}_\mathbf{1} \, , \\
\nonumber V^{\mathbf{8}(\mathbf{3})}_\mathbf{3} &=& 2
S^{\mathbf{8}(\mathbf{8})}_\mathbf{1} -
V^{\mathbf{8}(\mathbf{8})}_\mathbf{1} +
A^{\mathbf{8}(\mathbf{8})}_\mathbf{1} - 2
P^{\mathbf{8}(\mathbf{8})}_\mathbf{1} \, ,
\\ T^{\mathbf{8}(\mathbf{3})}_\mathbf{6} &=&
6 S^{\mathbf{8}(\mathbf{8})}_\mathbf{1} +
T^{\mathbf{8}(\mathbf{8})}_\mathbf{1} + 6
P^{\mathbf{8}(\mathbf{8})}_\mathbf{1} \, ,
\\ \nonumber A^{\mathbf{8}(\mathbf{3})}_\mathbf{6} &=&
2 S^{\mathbf{8}(\mathbf{8})}_\mathbf{1} +
V^{\mathbf{8}(\mathbf{8})}_\mathbf{1} -
A^{\mathbf{8}(\mathbf{8})}_\mathbf{1} - 2
P^{\mathbf{8}(\mathbf{8})}_\mathbf{1} \, ,
\\ \nonumber P^{\mathbf{8}(\mathbf{3})}_\mathbf{3} &=&
- \frac{1}{2} S^{\mathbf{8}(\mathbf{8})}_\mathbf{1} + \frac{1}{2}
V^{\mathbf{8}(\mathbf{8})}_\mathbf{1} + \frac{1}{4}
T^{\mathbf{8}(\mathbf{8})}_\mathbf{1} + \frac{1}{2}
A^{\mathbf{8}(\mathbf{8})}_\mathbf{1} - \frac{1}{2}
P^{\mathbf{8}(\mathbf{8})}_\mathbf{1}\, .
\end{eqnarray}
%

%
\subsection{Specifying the color structure}\label{app_color}
%

For completeness of mathematical structure, one can specify the
color quantum numbers for the currents rather the flavor ones. For
illustration, let us consider the color structure $(\mathbf{3}
\otimes \mathbf{3}) \otimes (\mathbf{\bar 3} \otimes \mathbf{\bar
3}) \rightarrow \mathbf{\bar 3} \otimes \mathbf{3} \rightarrow
\mathbf{1_c}$. In order to establish the relations between
$(qq)(\bar q \bar q)$ and $(\bar q q)(\bar q q)$ currents, we find
that we need two flavor structures: $(\mathbf{3_f} \otimes
\mathbf{3_f}) \otimes (\mathbf{\bar 3_f} \otimes \mathbf{\bar 3_f})
\rightarrow \mathbf{\bar 3_f} \otimes \mathbf{3_f} \rightarrow
\mathbf{1_f}$ and $(\mathbf{3_f} \otimes \mathbf{3_f}) \otimes
(\mathbf{\bar 3_f} \otimes \mathbf{\bar 3_f}) \rightarrow
\mathbf{6_f} \otimes \mathbf{\bar 6_f} \rightarrow \mathbf{1_f}$.

In this case, diquarks and antidiquarks have an anti-symmetric color
structure. By using the Pauli principle, we can verify
%
\begin{equation}
S^{\mathbf{1}(\mathbf{6})}_\mathbf{3} =
V^{\mathbf{1}(\mathbf{6})}_\mathbf{3} =
T^{\mathbf{1}(\mathbf{3})}_\mathbf{3} =
A^{\mathbf{1}(\mathbf{3})}_\mathbf{3} =
P^{\mathbf{1}(\mathbf{6})}_\mathbf{3} = 0 \, .
\end{equation}
%
Therefore, there are five types of $(qq)(\bar q \bar q)$ fields,
which are non-zero and independent:
%
\begin{equation} \nonumber
S^{\mathbf{1}(\mathbf{3})}_\mathbf{3} \, ,
V^{\mathbf{1}(\mathbf{3})}_\mathbf{3} \, ,
T^{\mathbf{1}(\mathbf{6})}_\mathbf{3} \, ,
A^{\mathbf{1}(\mathbf{6})}_\mathbf{3} \, ,
P^{\mathbf{1}(\mathbf{3})}_\mathbf{3} \, .
\end{equation}
%
The single $(\bar q q)(\bar q q)$ fields can not have an
anti-symmetric color structure. Therefore, we need to use their
combinations. By using Eq.~(\ref{eq_color}), $(\bar q q)(\bar q q)$
fields can be combined to have an anti-symmetric color structure:
%
\begin{eqnarray} \nonumber
(\bar q^A_a q^A_a)(\bar q^B_b q^B_b) - (\bar q^A_a q^A_b)(\bar q^B_b
q^B_a) &=& (\bar q^A_a q^A_a)(\bar q^B_b q^B_b) - {1 \over 3}(\bar
q^A_a q^A_a)(\bar q^B_b q^B_b) - {1 \over 2} \lambda_n^{ab}
\lambda_n^{cd} (\bar q^A_a q^A_b)(\bar q^B_c q^B_d) \\ &=& {2 \over
3} S^{\mathbf{1}(\mathbf{1})}_\mathbf{1} - {1 \over 2}
S^{\mathbf{1}(\mathbf{1})}_\mathbf{8} \equiv
S^{\mathbf{1}(\mathbf{1})}_\mathbf{3} \, ,
\end{eqnarray}
%
Altogether there are ten types of non-vanishing $(\bar qq)(\bar q
q)$ currents:
%
\begin{eqnarray} \nonumber &&
S^{\mathbf{1}(\mathbf{1})}_\mathbf{3} \, ,
V^{\mathbf{1}(\mathbf{1})}_\mathbf{3} \, ,
T^{\mathbf{1}(\mathbf{1})}_\mathbf{3} \, ,
A^{\mathbf{1}(\mathbf{1})}_\mathbf{3} \, ,
P^{\mathbf{1}(\mathbf{1})}_\mathbf{3} \, ,
S^{\mathbf{1}(\mathbf{8})}_\mathbf{3} \, ,
V^{\mathbf{1}(\mathbf{8})}_\mathbf{3} \, ,
T^{\mathbf{1}(\mathbf{8})}_\mathbf{3} \, ,
A^{\mathbf{1}(\mathbf{8})}_\mathbf{3} \, ,
P^{\mathbf{1}(\mathbf{8})}_\mathbf{3} \, .
\end{eqnarray}
%
Once again, among them only five are independent
%
\begin{eqnarray}\nonumber
S^{\mathbf{1}(\mathbf{8})}_\mathbf{3} &=&  -\frac{1}{6}
S^{\mathbf{1}(\mathbf{1})}_\mathbf{3} + \frac{1}{2}
V^{\mathbf{1}(\mathbf{1})}_\mathbf{3} + \frac{1}{4}
T^{\mathbf{1}(\mathbf{1})}_\mathbf{3} - \frac{1}{2}
A^{\mathbf{1}(\mathbf{1})}_\mathbf{3} - \frac{1}{2}
P^{\mathbf{1}(\mathbf{1})}_\mathbf{3}\, ,
\\ \nonumber V^{\mathbf{1}(\mathbf{8})}_\mathbf{3} &=& 2
S^{\mathbf{1}(\mathbf{1})}_\mathbf{3} - \frac{5}{3}
V^{\mathbf{1}(\mathbf{1})}_\mathbf{3} -
A^{\mathbf{1}(\mathbf{1})}_\mathbf{3} - 2
P^{\mathbf{1}(\mathbf{1})}_\mathbf{3}\, ,
\\ T^{\mathbf{1}(\mathbf{8})}_\mathbf{3} &=& 6
S^{\mathbf{1}(\mathbf{1})}_\mathbf{3} - \frac{5}{3}
T^{\mathbf{1}(\mathbf{1})}_\mathbf{3} + 6
P^{\mathbf{1}(\mathbf{1})}_\mathbf{3}\, ,
\\ \nonumber A^{\mathbf{1}(\mathbf{8})}_\mathbf{3} &=& - 2
S^{\mathbf{1}(\mathbf{1})}_\mathbf{3} -
V^{\mathbf{1}(\mathbf{1})}_\mathbf{3} - \frac{5}{3}
A^{\mathbf{1}(\mathbf{1})}_\mathbf{3} + 2
P^{\mathbf{1}(\mathbf{1})}_\mathbf{3}\, ,
\\ \nonumber P^{\mathbf{1}(\mathbf{8})}_\mathbf{3} &=& \frac{1}{2}
S^{\mathbf{1}(\mathbf{1})}_\mathbf{3} - \frac{1}{2}
V^{\mathbf{1}(\mathbf{1})}_\mathbf{3} + \frac{1}{4}
T^{\mathbf{1}(\mathbf{1})}_\mathbf{3} + \frac{1}{2}
A^{\mathbf{1}(\mathbf{1})}_\mathbf{3} - \frac{1}{6}
P^{\mathbf{1}(\mathbf{1})}_\mathbf{3}\, .
\end{eqnarray}
%
The relations between $(qq)(\bar q \bar q)$ and $(\bar q q)(\bar q
q)$ structures are:
%
\begin{eqnarray}
\nonumber S^{\mathbf{1}(\mathbf{3})}_\mathbf{3} &=& -\frac{1}{2}
S^{\mathbf{1}(\mathbf{1})}_\mathbf{3} - \frac{1}{2}
V^{\mathbf{1}(\mathbf{1})}_\mathbf{3} + \frac{1}{4}
T^{\mathbf{1}(\mathbf{1})}_\mathbf{3}
- \frac{1}{2} A^{\mathbf{1}(\mathbf{1})}_\mathbf{3} - \frac{1}{2} P^{\mathbf{1}(\mathbf{1})}_\mathbf{3} \, , \\
\nonumber V^{\mathbf{1}(\mathbf{3})}_\mathbf{3} &=& 2
S^{\mathbf{1}(\mathbf{1})}_\mathbf{3} -
V^{\mathbf{1}(\mathbf{1})}_\mathbf{3} +
A^{\mathbf{1}(\mathbf{1})}_\mathbf{3} - 2
P^{\mathbf{1}(\mathbf{1})}_\mathbf{3} \, ,
\\ T^{\mathbf{1}(\mathbf{6})}_\mathbf{3} &=& 6 S^{\mathbf{1}(\mathbf{1})}_\mathbf{3}
+ T^{\mathbf{1}(\mathbf{1})}_\mathbf{3} + 6
P^{\mathbf{1}(\mathbf{1})}_\mathbf{3} \, ,
\\ \nonumber A^{\mathbf{1}(\mathbf{6})}_\mathbf{3} &=& 2 S^{\mathbf{1}(\mathbf{1})}_\mathbf{3}
+ V^{\mathbf{1}(\mathbf{1})}_\mathbf{3} -
A^{\mathbf{1}(\mathbf{1})}_\mathbf{3} - 2
P^{\mathbf{1}(\mathbf{1})}_\mathbf{3} \, ,
\\ \nonumber P^{\mathbf{1}(\mathbf{3})}_\mathbf{3} &=& -\frac{1}{2} S^{\mathbf{1}(\mathbf{1})}_\mathbf{3}
+ \frac{1}{2} V^{\mathbf{1}(\mathbf{1})}_\mathbf{3} + \frac{1}{4}
T^{\mathbf{1}(\mathbf{1})}_\mathbf{3} + \frac{1}{2}
A^{\mathbf{1}(\mathbf{1})}_\mathbf{3} - \frac{1}{2}
P^{\mathbf{1}(\mathbf{1})}_\mathbf{3}\, .
\end{eqnarray}
%

%
\subsection{Specifying the Lorentz structure}\label{app_lorentz}
%

Finally, let us consider the case where the Lorentz structure is
specified. As an illustration, let us consider a tetraquark current
$(q^T C \gamma_5 q)(\bar q \gamma_5 C \bar q^T)$. Possible color
structures are $(\mathbf{3} \otimes \mathbf{3}) \otimes
(\mathbf{\bar 3} \otimes \mathbf{\bar 3}) \rightarrow \mathbf{\bar
3} \otimes \mathbf{3} \rightarrow \mathbf{1_c}$ and $(\mathbf{3}
\otimes \mathbf{3}) \otimes (\mathbf{\bar 3} \otimes \mathbf{\bar
3}) \rightarrow \mathbf{6} \otimes \mathbf{\bar 6} \rightarrow
\mathbf{1_c}$; and possible flavor structures are $(\mathbf{3}
\otimes \mathbf{3}) \otimes (\mathbf{\bar 3} \otimes \mathbf{\bar
3}) \rightarrow \mathbf{\bar 3} \otimes \mathbf{3} \rightarrow
\mathbf{1_f}$ and $(\mathbf{3} \otimes \mathbf{3}) \otimes
(\mathbf{\bar 3} \otimes \mathbf{\bar 3}) \rightarrow \mathbf{6}
\otimes \mathbf{\bar 6} \rightarrow \mathbf{1_f}$.

By using the Pauli principle, we can verify
%
\begin{equation}
S^{\mathbf{1}(\mathbf{6})}_\mathbf{3} =
S^{\mathbf{1}(\mathbf{3})}_\mathbf{6} = 0 \, .
\end{equation}
%
Therefore, there are two currents which are non-zero and
independent:
%
\begin{eqnarray} \nonumber
S^{\mathbf{1}(\mathbf{3})}_\mathbf{3} &=&
\epsilon^{abe}\epsilon^{cde}\epsilon_{ABE}\epsilon_{CDE} (q^A_a C
\gamma_5 q^B_b)(\bar q^C_c \gamma_5 C \bar q^D_d) \, ,
\\ \nonumber
S^{\mathbf{1}(\mathbf{6})}_\mathbf{6} &=& (q^A_a C \gamma_5
q^B_b)(\bar q^A_a \gamma_5 C \bar q^B_b + \bar q^B_a \gamma_5 C \bar
q^A_b + (a \leftrightarrow b)) \, ,
\end{eqnarray}
%
Now from the combination of quark and antiquark, possible color
structures are $(\mathbf{\bar 3} \otimes \mathbf{3}) \otimes
(\mathbf{\bar 3} \otimes \mathbf{3}) \rightarrow \mathbf{1} \otimes
\mathbf{1} \rightarrow \mathbf{1_c}$ and $(\mathbf{\bar 3} \otimes
\mathbf{3}) \otimes (\mathbf{\bar 3} \otimes \mathbf{3}) \rightarrow
\mathbf{8} \otimes \mathbf{8} \rightarrow \mathbf{1_c}$; and
possible flavor structures are $(\mathbf{\bar 3} \otimes \mathbf{3})
\otimes (\mathbf{\bar 3} \otimes \mathbf{3}) \rightarrow \mathbf{1}
\otimes \mathbf{1} \rightarrow \mathbf{1_f}$ and $(\mathbf{\bar 3}
\otimes \mathbf{3}) \otimes (\mathbf{\bar 3} \otimes \mathbf{3})
\rightarrow \mathbf{8} \otimes \mathbf{8} \rightarrow \mathbf{1_f}$.
Therefore, there are four non-vanishing currents:
%
\begin{eqnarray} \nonumber
P^{\prime\mathbf{1}(\mathbf{1})}_\mathbf{1} &=& (q^A_a C \gamma_5
q^B_b)(\bar q^A_a \gamma_5 C \bar q^B_b) \, ,
\\ \nonumber P^{\prime\mathbf{1}(\mathbf{1})}_\mathbf{8} &=& \lambda_n^{ab}\lambda_n^{cd}
(q^A_a C \gamma_5 q^B_c)(\bar q^A_b \gamma_5 C \bar q^B_d) \, ,
\\ \nonumber P^{\prime\mathbf{1}(\mathbf{8})}_\mathbf{1} &=& \lambda^N_{AB}\lambda^N_{CD}
(q^A_a C \gamma_5 q^C_b)(\bar q^B_a \gamma_5 C \bar q^D_b) \, ,
\\ \nonumber P^{\prime\mathbf{1}(\mathbf{8})}_\mathbf{8} &=& \lambda^N_{AB}\lambda^N_{CD}
\lambda_n^{ab}\lambda_n^{cd} (q^A_a C \gamma_5q^C_c)(\bar q^B_b
\gamma_5 C \bar q^D_d) \, .
\end{eqnarray}
%
The Lorentz structure is still specified to be $(q^T C \gamma_5
q)(\bar q \gamma_5 C \bar q^T)$. However, if we interchange the
second quark and third antiquark as done in Eq.~(\ref{ex_cf}) within
the color and flavor spaces structures, They are now ``$(\bar q
q)(\bar q q)$'' currents. Among them, only two are independent,
through the following relations:
%
\begin{eqnarray} \nonumber
P^{\prime\mathbf{1}(\mathbf{8})}_\mathbf{1} &=&
P^{\prime\mathbf{1}(\mathbf{1})}_\mathbf{8} \, ,
\\ P^{\prime\mathbf{1}(\mathbf{8})}_\mathbf{8} &=&
{32 \over 9} P^{\prime\mathbf{1}(\mathbf{1})}_\mathbf{1} - {4 \over
3} P^{\prime\mathbf{1}(\mathbf{1})}_\mathbf{8} \, .
\end{eqnarray}
%
Finally, relations between the $(q q)(\bar q \bar q)$ and ``$(\bar q
q)(\bar q q)$'' currents are
%
\begin{eqnarray}\nonumber
S^{\mathbf{1}(\mathbf{3})}_{\mathbf{3}} &=& {4 \over 3}
P^{\prime\mathbf{1}(\mathbf{1})}_{\mathbf{1}} -
P^{\prime\mathbf{1}(\mathbf{1})}_{\mathbf{8}} \, ,
\\ S^{\mathbf{1}(\mathbf{6})}_{\mathbf{6}} &=& {8 \over 3}
P^{\prime\mathbf{1}(\mathbf{1})}_{\mathbf{1}} +
P^{\prime\mathbf{1}(\mathbf{1})}_{\mathbf{8}} \, .
\end{eqnarray}
%

%

%

\end{document}